\documentclass[twocolumn,prb,eqsecnum,showpacs,preprintnumbers,amsmath,amssymb]{revtex4}

\usepackage{graphicx}% Include figure files
\usepackage{dcolumn}% Align table columns on decimal point
\usepackage{bm}% bold math
\include{pdfsync}

\begin{document}
\title{Energy spectrum and broken spin-surface 
locking in topological insulator quantum dots}
\author{Arijit Kundu,$^1$ Alex Zazunov,$^1$ Alfredo Levy Yeyati,$^2$ Thierry
Martin,$^{3,4}$ and Reinhold Egger$^1$ }
\affiliation{$^1$~Institut f\"ur Theoretische Physik,
Heinrich-Heine-Universit\"at, D-40225 D\"usseldorf, Germany\\
$^2$Departamento de F{\'i}sica Te{\'o}rica de la Materia Condensada C-V,
Universidad Aut{\'o}noma de Madrid, E-28049 Madrid, Spain \\
$^3$~Centre de Physique Th{\'e}orique, Campus de Luminy, 
case 907, F-13288 Marseille, France\\
$^4$Universit{\'e} de la M{\'e}dit{\'e}rann{\'e}e,  F-13288 Marseille, France}

\date{\today}
\begin{abstract}
We consider the energy spectrum and the spin-parity structure of
the eigenstates for a quantum dot made of a strong topological insulator.  
Using the effective low-energy theory in a finite-length
cylinder geometry,  numerical calculations show that even at the lowest
energy scales, the spin direction in a topologically protected 
surface mode is not locked to the surface.  
We find ``zero-momentum''
modes, and subgap states localized near the ``caps'' of the dot.
Both the energy spectrum and the spin texture of the eigenstates
are basically reproduced from an analytical surface Dirac fermion description.  
Our results are compared to microscopic calculations using
a tight-binding model for a strong topological insulator in a finite-length
nanowire geometry.
\end{abstract}


\maketitle
\section{Introduction}

The theoretical prediction and subsequent experimental verification
of the conducting surface state of a strong 
topological insulator (TI) continues to generate a lot of 
excitement in physics; for reviews, see Refs.~\onlinecite{hasan,qi,joel}. 
In a TI, strong spin-orbit couplings and band inversion 
conspire to produce a time-reversal invariant topological state 
different from a conventional band insulator.
Using Bi$_2$Se$_3$ as a weakly correlated 
reference TI material with rather large bulk 
gap $\Delta_b \approx 0.3$~meV,  surface probe experiments 
have provided strong evidence for the topologically protected 
gapless surface state.\cite{xia} The measured spin texture of the surface 
state is consistent with predictions obtained for 
two-dimensional (2D) massless Dirac fermions.  
Under this ``relativistic'' description, the spin direction is 
locked to the surface, and the surface state is stable 
against the effects of weak disorder and
weak interactions (topological protection).\cite{hasan,qi,joel}
It is thus useful to first study the simplest case of a noninteracting 
disorder-free model, which is the case investigated below.

Because of the residual bulk conductivity of the  presently available
(nominally insulating) TI samples, 
it has been difficult to experimentally extract the surface contribution 
to the electrical conductivity. One attempt to improve the situation is 
to consider mesoscopic samples, where the surface-to-volume ratio is 
more advantageous.
In particular, thin-film geometries\cite{qu,checkelsky} and
quasi-1D nanowires (``ribbons'')\cite{peng,cha,zuev,tang}   have been studied
experimentally.  Signatures for Aharonov-Bohm 
interference effects associated with the topological 
surface state in Bi$_2$Se$_3$ nanowires  
were reported,\cite{peng} cf.~also related 
experiments for Sb$_2$Te$_3$ nanowires.\cite{zuev}
For infinitely long and circulary symmetric topological nanowires, 
band structure calculations predict a multi-channel waveguide 
where all surface modes are gapped because
of spin-surface locking.\cite{ashv0,egger,piet,ashvin} 
With $k$ denoting the conserved momentum along the wire axis (taken along
the $z$ direction),
and $j$ the half-integer total angular momentum, 
the dispersion relation of these modes is  
\begin{equation}\label{bulkdisp}
E_{j,\pm}(k)=\pm \sqrt{ (v_1 k)^2+  ( j v_2/R)^2 },
\end{equation}
where $\hbar=1$ throughout and $\pm$ for conduction and valence band, 
respectively.  The Fermi velocities $v_1$ and $v_2$ differ because
the bulk dispersion relation is anisotropic, see below, and the
nanowire radius is $R$.  Note that there is a minimal gap 
$\Delta_s= v_2/R$ for the surface modes since $j$ is half-integer.
For reasonable values of $R$, we have  $\Delta_s\ll \Delta_b$.

Experiments probing quantum dot physics in 
finite-length TI nanowires  are expected to yield new insights into the 
exciting physics of TIs, in close 
analogy to semiconductor nanowires and carbon nanotubes
where such experiments have been highly successful.\cite{nano1,nano2}
To prepare the ground, we here consider the 
band structure for TI quantum dots with a finite-length nanowire geometry.
For a spherical TI dot, the band structure was  worked out 
before.\cite{dhlee,paco,malkova}  
However, we draw attention to several features that only arise 
when the surface contains sharp edges, i.e., non-differentiable parts,
as is the case for the cylinder. 
We employ three different and 
independent approaches to understand TI quantum dot energy levels
and their spin texture:
(1) For a cylindrical TI nanowire of length $L$
and radius $R$ closed by flat caps, 
we have performed detailed numerical calculations for the energy 
spectrum and the spin texture of the eigenstates
based on the effective low-energy theory of 
Zhang \textit{et al.};\cite{zhang1,zhang2} 
material parameters were chosen for Bi$_2$Se$_3$ as quoted in
Ref.~\onlinecite{zhang2}.  
(2) An analytical approach starting from a surface Dirac fermion description
has been developed for the same geometry.  Most of our numerical results
can thereby be quantitatively reproduced within an analytical theory.
(3)  We have also studied a microscopic tight-binding model for
a strong TI in the finite-size nanowire geometry and find 
qualitatively similar results. 

The main conclusions reached from these three approaches are as follows:
First, longitudinal momentum quantization implies a discrete
sequence of energy levels which can (roughly) be approximated by 
letting $k\to k_n=n\pi /L$ ($n\in \mathbb{N}$) in Eq.~\eqref{bulkdisp}. 
Remarkably, in addition we find unconventional
 ``zero-momentum'' states (where, formally, $k_n=0$). 
In such a state, the charge and spin densities are almost 
homogeneous along the $z$-direction.
Furthermore, there are subgap states energetically
located within the surface mode gap ($\Delta_s$). These states
are localized near both caps and show interesting spin texture.
Second, we observe significant out-of-surface components for the spin density
associated with all energy eigenstates.
Note that an out-of-plane spin texture is only expected when trigonal warping 
effects are important,\cite{fu,yazyev} 
see also very recent experimental results reporting such 
features.\cite{trig1,trig2} 
However, to lowest order in 
momentum (around the $\Gamma$ point), trigonal warping can be neglected,
while we persistently find broken spin-surface locking also at the
lowest energy scales.  
Our observations are instead related to the presence of non-differentiable
sections of the surface, where the wavefunctions in the ``trunk'' and 
``cap'' regions of the cylinder have to be matched.  
Such non-differentiable surface parts also appear in the 
samples studied in Refs.~\onlinecite{peng,cha}, and 
therefore surface probe experiments for these devices
could directly test our prediction of broken spin-surface locking.
For differentiable closed surfaces, we expect spin-surface
locking to stay intact. In fact, the explicit solution of the problem in a 
spherical geometry exhibits spin-surface locking.\cite{dhlee,paco} 
We note in passing that for a flat TI surface, 
a time-dependent out-of-plane spin component 
can also be generated by elastic disorder.
However, this component will precess around the momentum-dependent
spin-orbit axis (which lies in the plane) and averages to 
zero on time scales corresponding to the inverse Fermi energy.\cite{burkov}

The structure of the remainder of this paper is as follows. 
In Sec.~\ref{sec2}, we describe the results of our numerical calculations for 
cylindrical TI nanowire dots based on the effective low-energy theory.
In Sec.~\ref{sec3}, an
analytical approach based on the surface Dirac fermion picture is described
for the same geometry.  The spectrum and the spin density 
profile for the resulting eigenstates will be derived, and the
results are compared to the numerical findings in Sec.~\ref{sec2}.  
In Sec.~\ref{sec4}, we turn to the  tight-binding calculation and 
compare those computations  to the previous results.  
Finally, we conclude in Sec.~\ref{sec5}.  Appendix \ref{appa}
contains a derivation of the surface Dirac fermion Hamiltonian 
for an infinitely long nanowire using the approach of Ref.~\onlinecite{paco}.

\section{Effective low-energy description} \label{sec2}

\subsection{Model and numerical approach}\label{sec2a}

In this section, we compute the band structure of a cylindrical nanowire
of length $L$ and radius $R$ from the effective low-energy theory
of Zhang \textit{et al.}\cite{zhang1,zhang2} using  parameters for Bi$_2$Se$_3$.
Up to terms of order ${\bm k}^2$ with ${\bm k}=(k_x,k_y,k)$, 
the low-energy bulk Hamiltonian employing the four bands 
energetically closest to the $\Gamma$ point is \cite{zhang2}
\begin{equation}\label{bulkham}
H_b =  \epsilon_{\bm k} \sigma_0\tau_0 + M_{\bm k} \sigma_0\tau_z +
[A_0 (k_x \sigma_x+k_y\sigma_y) + B_0 k\sigma_z]\tau_x,
\end{equation}
where $\epsilon_{\bm k}=C_0+C_1k^2+C_2 k_\perp^2$ and $M_{\bm k}=
M_0 + M_1 k^2+M_2 k_\perp^2$, with $k_\perp^2=k_x^2+k_y^2$. 
This also defines the Fermi velocities $v_1= B_0/\hbar$ (in $z$ direction) 
and $v_2= A_0/\hbar$ (in the $xy$ plane), see Eq.~\eqref{bulkdisp}.
The basis states encoding the spin-parity structure, where Pauli matrices
$\sigma_i$ ($\tau_i$) act in spin (parity) space and $\sigma_0$ ($\tau_0$) 
denotes the respective unit matrix, are explicitly
given in Ref.~\onlinecite{zhang2}.  This work also describes
the extension of Eq.~\eqref{bulkham} to the case of eight bands or towards
including trigonal warping.  Using the parameters in 
Ref.~\onlinecite{zhang2}, listed 
for convenience also in Table \ref{tab0}, the criterion for a TI 
phase\cite{hasan} is satisfied.  One therefore must have 
an odd number of conducting surface modes when boundaries are present.

\begin{widetext}
\begin{table}[t]
\begin{tabular}{|c|c|c|c|c|c|c|c|}
\hline
$A_0$ (eV\AA)  & $B_0$ (eV\AA) & $C_0$ (eV)   & $C_1$
(eV\AA$^2$) & $C_2$ (eV\AA$^2$)& $M_0$ (eV) &  $M_1$ (eV\AA$^2$) & $M_2$
(eV\AA$^2$) \\ \hline
3.33   & 2.26  & -0.0083 & 5.74  & 30.4  &  -0.28 & 6.86 & 44.5 \\ 
\hline
\end{tabular}
\caption{\label{tab0} 
Parameter values in Eq.~\eqref{bulkham} appropriate 
for Bi$_2$Se$_3$  (taken from Ref.~\onlinecite{zhang2}).}
\end{table}
\end{widetext}

Our aim is to describe the band structure of a finite-length 
cylindrical nanowire with axis along the $z$ direction.
Due to rotational symmetry in the $xy$ plane, it is 
useful to switch to cylindrical coordinates $(r, \phi, z)$.
The ``cylindrical'' Pauli matrices ($\sigma_{r},\sigma_\phi,\sigma_z$) 
then represent the physical spin operator,\cite{zhang1,zhang2} 
\begin{equation}\label{cylpaul}
\sigma_{r,\phi} = e^{-i\sigma_z\phi/2} \sigma_{x,y} e^{i\sigma_z\phi/2},
\end{equation}  
and we refer to their local expectation values as ``spin densities'' below.
The conserved total angular momentum operator is
\begin{equation}\label{jdef}
\hat J = e^{-i \sigma_z \phi/2} 
\left( -i \partial_\phi \right) e^{i \sigma_z \phi/2} 
=-i \partial_\phi + \sigma_z / 2 .
\end{equation}
For the finite-length cylinder we then construct the eigenfunctions
to the Hamiltonian \eqref{bulkham} with 
Dirichlet boundary conditions, $\Psi({\bm r})=0$, on the
surface, i.e., for $|z|<L/2$ with $r=R$ (cylinder trunk)
and for $|z|=L/2$ with $r<R$ (caps).  This is automatically achieved by 
expanding states in a complete orthonormal basis,
$\{ \psi_a (r,\phi,z) \}$,
that satisfies these boundary conditions. The quantum numbers 
$a=(j,\nu,n,\sigma)$ include the half-integer angular momentum $j$, 
a radial index $\nu\in \mathbb{N}$, the longitudinal quantum
number $n\in \mathbb{N}$, and the spin index $\sigma=\pm$.  
Explicitly, see also Ref.~\onlinecite{egger},
for $r\le R$ and $|z|\le L/2$, the basis is chosen in the form
\begin{equation}\label{ortho}
\psi_{a}(r,\phi,z) = \sqrt{\frac{2}{V}}\sin[ \pi n (z/L-1/2) ] e^{im\phi}
\frac{J_m(\gamma_{m\nu} r/R) }{J_{m+1}(\gamma_{m\nu})} ,
\end{equation}
where $m\equiv j-\sigma/2$, $V=\pi R^2 L$ is the cylinder volume, and 
$\gamma_{m\nu}$ denotes the $\nu$th zero of the Bessel function $J_m$.
The basis set \eqref{ortho} satisfies the orthonormality relation 
$\int_V d^3 {\bm r} \ \psi_a^\ast({\bm r})\psi_{a'}({\bm r}) = \delta_{aa'}.$ 
In addition, the basis states acquire a spinor structure in 
parity space not shown explicitly in Eq.~\eqref{ortho}.

Expanding the Hamiltonian $H_b$ [Eq.~\eqref{bulkham}] in this basis,
we obtain a matrix representation that allows for 
numerical calculations in a truncated basis set.  
Upon increasing the basis set, numerical results for the spectrum 
turn out to converge rather slowly. We have performed a 
lattice regularization as in Ref.~\onlinecite{koenig} in order to 
obtain manageable matrix dimensions.
Typically, we achieve convergence with $\approx 8000$ basis states
for given $j$.  The solution of the eigenvalue problem then
yields the discrete energy spectrum of such a quantum dot, 
$E=E_{j,s,\pm}$, where $s\in \mathbb{N}$
labels the different states for the conduction or valence ($\pm$) band 
with given angular momentum $j$.  Taking averages with respect
to the corresponding eigenvector
 $|\Psi_{j,s,\pm}\rangle$ then yields the 
spatially dependent charge density profile for this state, 
$\langle \rho \rangle(r,z)$.  In addition, one obtains 
the local spin densities, $\langle \sigma_{\alpha}\rangle(r,z)$ with 
$\alpha=r,\phi,z$, and the local parity  densities,
$\langle \tau_{\beta}\rangle (r,z)$ with $\beta=x,y,z$. 
Rotational symmetry implies that all these averages are independent 
of the angular variable $\phi$.  

\subsection{Numerical results}\label{sec2b}

\begin{figure}
\includegraphics[width=0.45\textwidth]{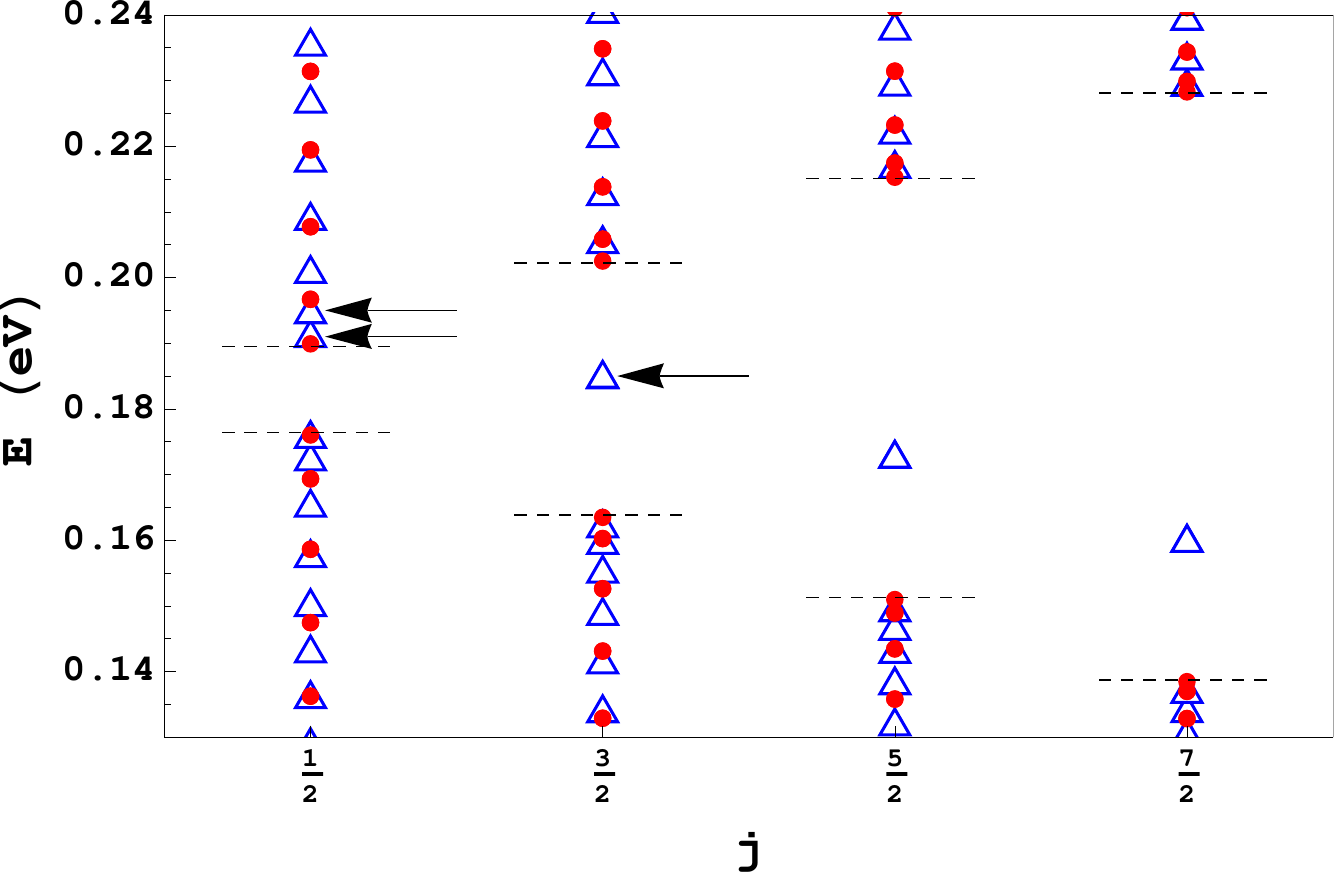}
\caption{\label{fig1}  (Color online)  Energy spectrum $E_{j,s,\pm}$ from 
numerical diagonalization of Eq.~\eqref{bulkham} for a cylindrical TI quantum 
dot with $R=20$~nm and $L=44$~nm.  Open blue triangles correspond to
the numerical results, while filled red circles 
show the analytical prediction in Eq.~\eqref{mom}. Dashed lines
indicate the surface gap for the respective angular momentum ($j$) mode in 
an infinitely long nanowire.  The spin texture for the zero-momentum
state with $j=1/2$ indicated by the lower
left arrow is shown in Fig.~\ref{fig2}.  The spin texture
for the next higher state (higher left arrow) is shown in Fig.~\ref{fig3}.
The example of a subgap state (for $j=3/2$, right arrow) is addressed
in Fig.~\ref{fig4}. 
}
\end{figure}

We now present the results of our numerical calculations.
The discrete energy spectrum for a TI nanowire dot with 
$R=20$~nm and $L=44$~nm  is shown in Fig.~\ref{fig1}.
The Kramers degeneracy results in an identical spectrum for $j\to -j$
but with reversed spin and parity ($\tau_y$) directions. We therefore
show only the $j>0$ solutions in Fig.~\ref{fig1}.  Moreover,
we focus on the topologically protected 
surface fermion modes inside the bulk gap $\Delta_b$.

There are several noteworthy points about Fig.~\ref{fig1}. First,
comparison with our analytical results,
see Eq.~\eqref{mom} and Sec.~\ref{sec3} below, shows that most levels
are approximately recovered from the 
bulk dispersion relation [Eq.~\eqref{bulkdisp}]
by simply imposing the standard quantization condition $k_n=n\pi/L$
with $n\in \mathbb{N}$ on the longitudinal momentum $k$. 
However, here additional states corresponding to $n=0$ emerge.
These \textit{zero-momentum states} are absent 
for Schr\"odinger fermions in a box.  Note that for each $j$,
there is precisely one $n=0$ state for
the conduction band and one for the valence band. 
Inspection of the density profiles for these states
reveals almost homogeneous charge, spin,
and parity densities as a function of the $z$ coordinate.
Second, for $j>1/2$, we find a pair of nearly degenerate
\textit{subgap states} inside the surface gap $\Delta_s$. 
(The near-degeneracy is not visible in Fig.~\ref{fig1}.)
Such a subgap state is localized with equal occupation 
probability  at \textit{both}\ caps.
Furthermore, electron-hole symmetry is broken under the Zhang model $H_b$,
in contrast to the analytical model in Sec.~\ref{sec3}. This is the 
main reason for the existing discrepancies between
Eq.~(\ref{mom})  and the numerical results, see Fig.~\ref{fig1}.  
In fact, we have also carried out additional numerical calculations
for an electron-hole symmetric version of Eq.~\eqref{bulkham}, where the
corresponding results fit almost perfectly to the analytical results
in Sec.~\ref{sec3}.  In particular, all subgap states then disappear.

\begin{figure}
\includegraphics[width=0.3\textwidth,angle=-90]{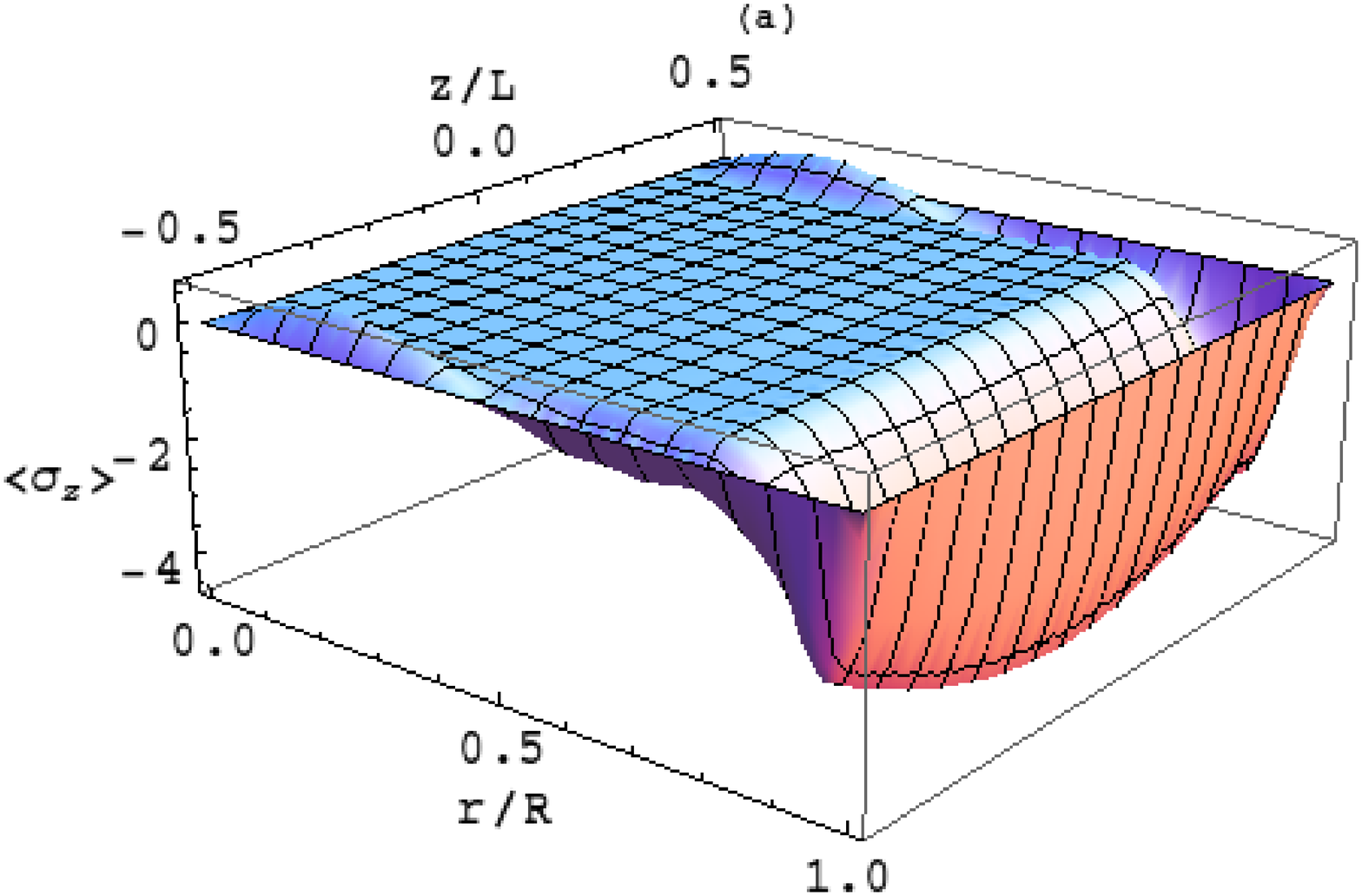}
\includegraphics[width=0.3\textwidth,angle=-90]{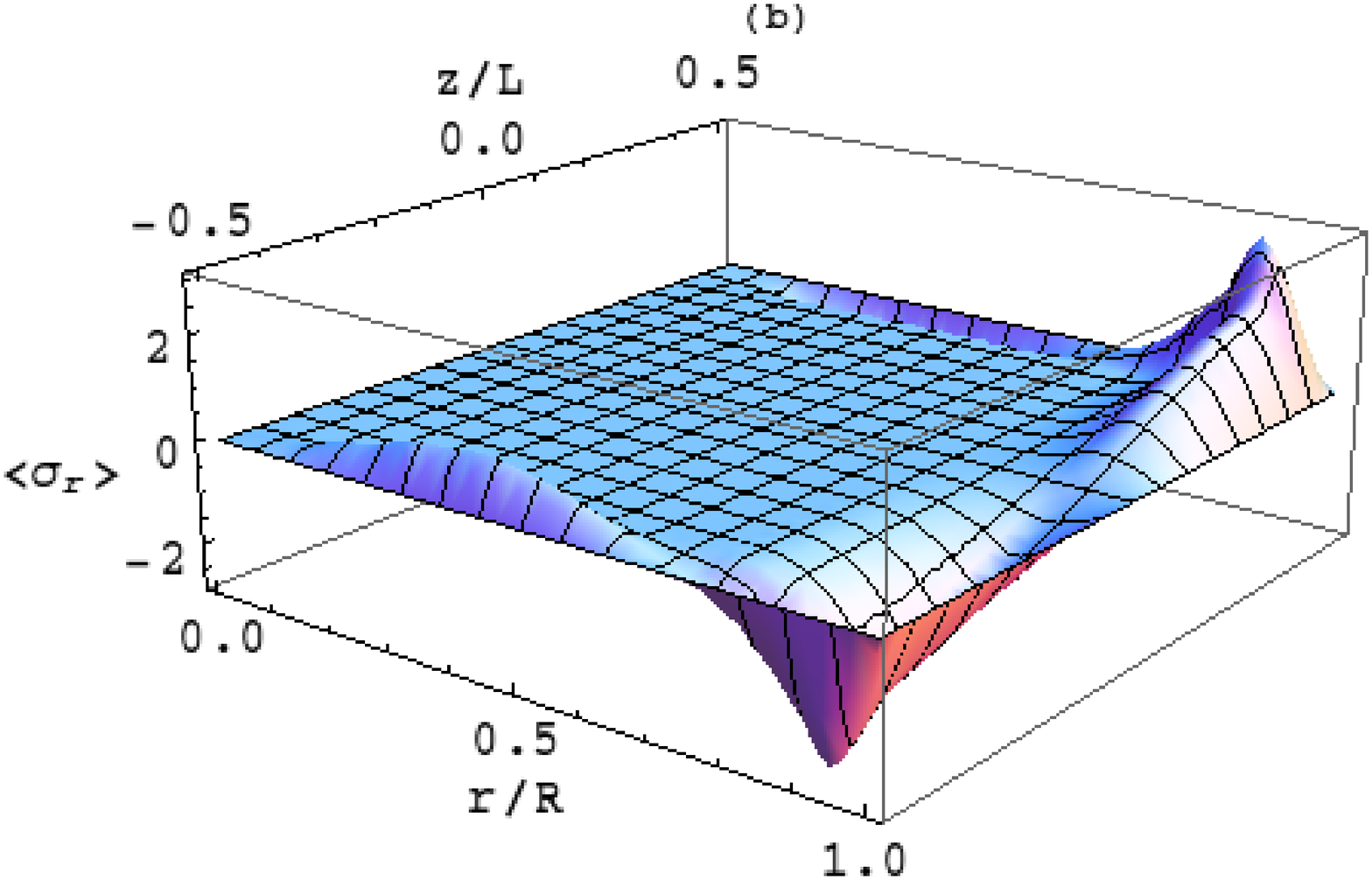}
\caption{\label{fig2}  (Color online) 
Spin density components $\langle\sigma_z\rangle$  [(a)]
and $\langle\sigma_r\rangle$ [(b)]
in the $rz$ plane, for the zero-momentum state with $j=1/2$
indicated by the lower left arrow in Fig.~\ref{fig1}.  }
\end{figure}

Inspection of the spin densities, $\langle \sigma_\alpha\rangle(r,z)$,
and parity densities, $\langle \tau_\beta \rangle(r,z)$,
for a given eigenstate $(j,s,\pm)$ yields 
\begin{equation} \label{numzero}
\langle \sigma_\phi \rangle(r,z) = \langle \tau_x \rangle(r,z) = 0 ,
\end{equation}
i.e., spin is never oriented in the circumferential direction.
In addition, there is now a finite radial spin component
within the trunk region ($|z|<L/2$)
and a finite $z$-component within the caps ($|z|=L/2$).
Hence, in general, the spin direction for a surface state
points out of the surface: \textit{spin-surface locking
is broken} in this geometry.  
This finding is in striking contrast to what happens in an
infinite nanowire\cite{egger} and for a sphere.\cite{dhlee}
Specifically, in the infinite cylinder case, the 
results corresponding to Fig.~\ref{fig2} show 
 $\langle\sigma_r\rangle=0$ reflecting spin-surface locking.

\begin{figure}
\includegraphics[width=0.3\textwidth,angle=-90]{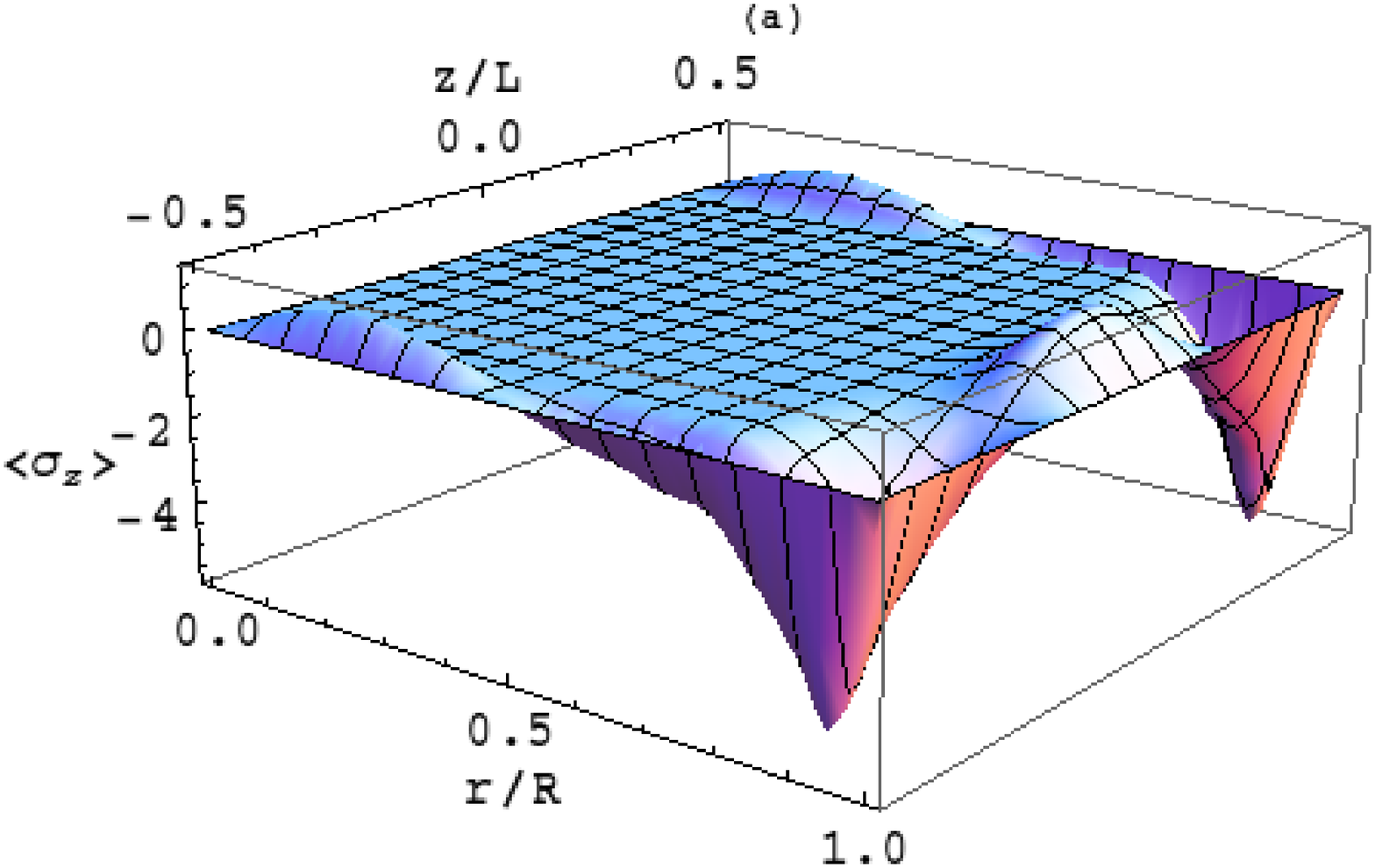}
\includegraphics[width=0.3\textwidth,angle=-90]{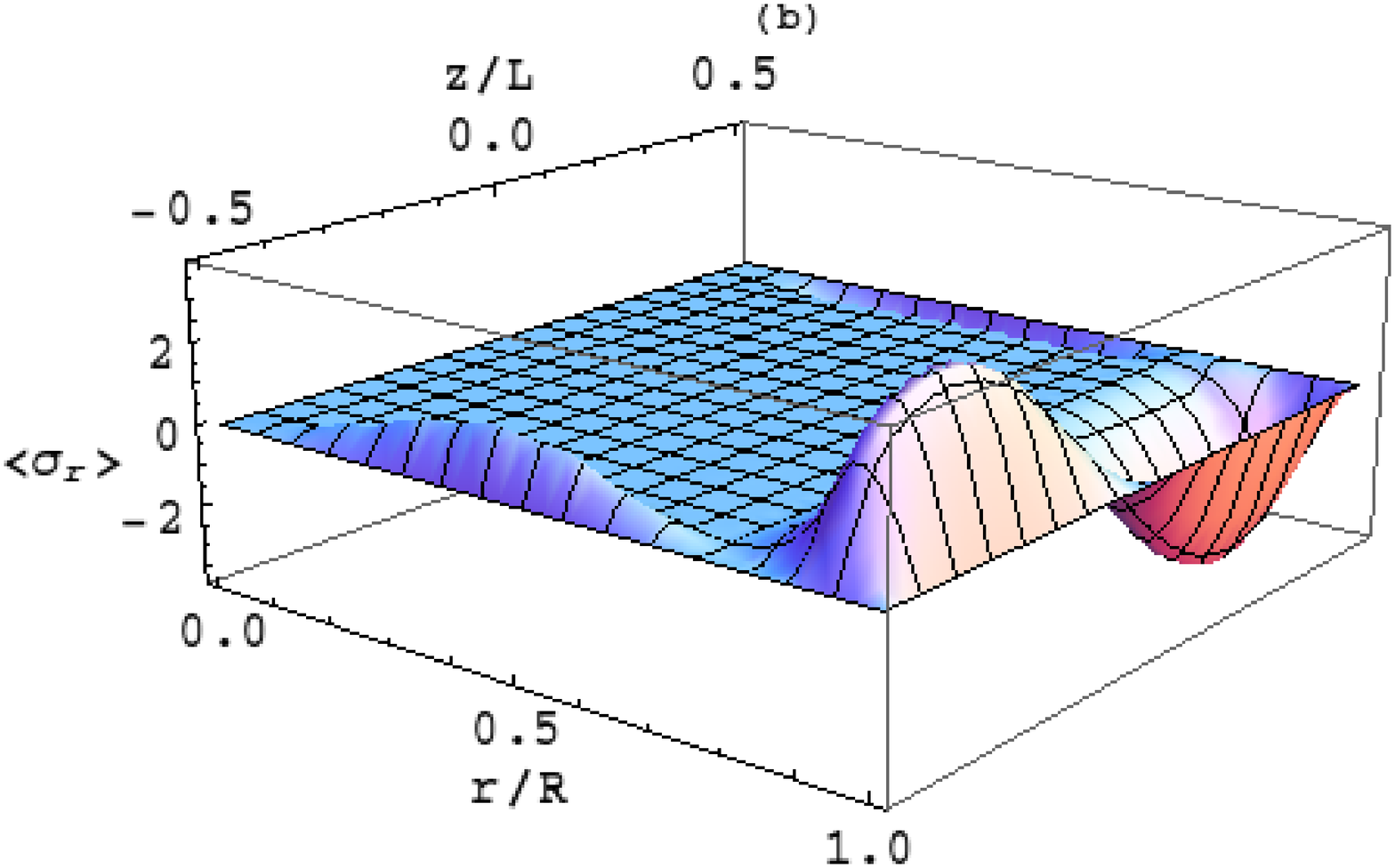}
\caption{\label{fig3}  (Color online) 
Same as Fig.~\ref{fig2} but for the next higher energy level 
(upper left arrow in Fig.~\ref{fig1}).
}
\end{figure}

Figure \ref{fig2} shows the spin density profile for the lowest-lying
(``zero-momentum'') conduction band state with $j=1/2$,
indicated by the lower left arrow in Fig.~\ref{fig1}.  
We indeed find an almost homogeneous spin density profile along the trunk,
where spin is mostly aligned along the (negative) $z$-direction, see
Fig.~\ref{fig2}(a).
Also the charge density is practically homogeneous along the 
$z$-direction (data not shown).
However, there is also a finite radial spin component breaking
spin-surface locking, see Fig.~\ref{fig2}(b).  For the cap region,
spin is mostly aligned along the radial direction, but again
an out-of-plane component, now oriented along the $z$-axis, is 
clearly visible.  For comparison, Fig.~\ref{fig3} shows the respective
results for the next higher energy 
level (upper left arrow in Fig.~\ref{fig1}).
Again we observe that spin-surface locking is violated, while in the
infinite cylinder case one finds $\langle \sigma_r = 0\rangle$ (spin-surface
locking).

\begin{figure}
\includegraphics[width=0.3\textwidth,angle=-90]{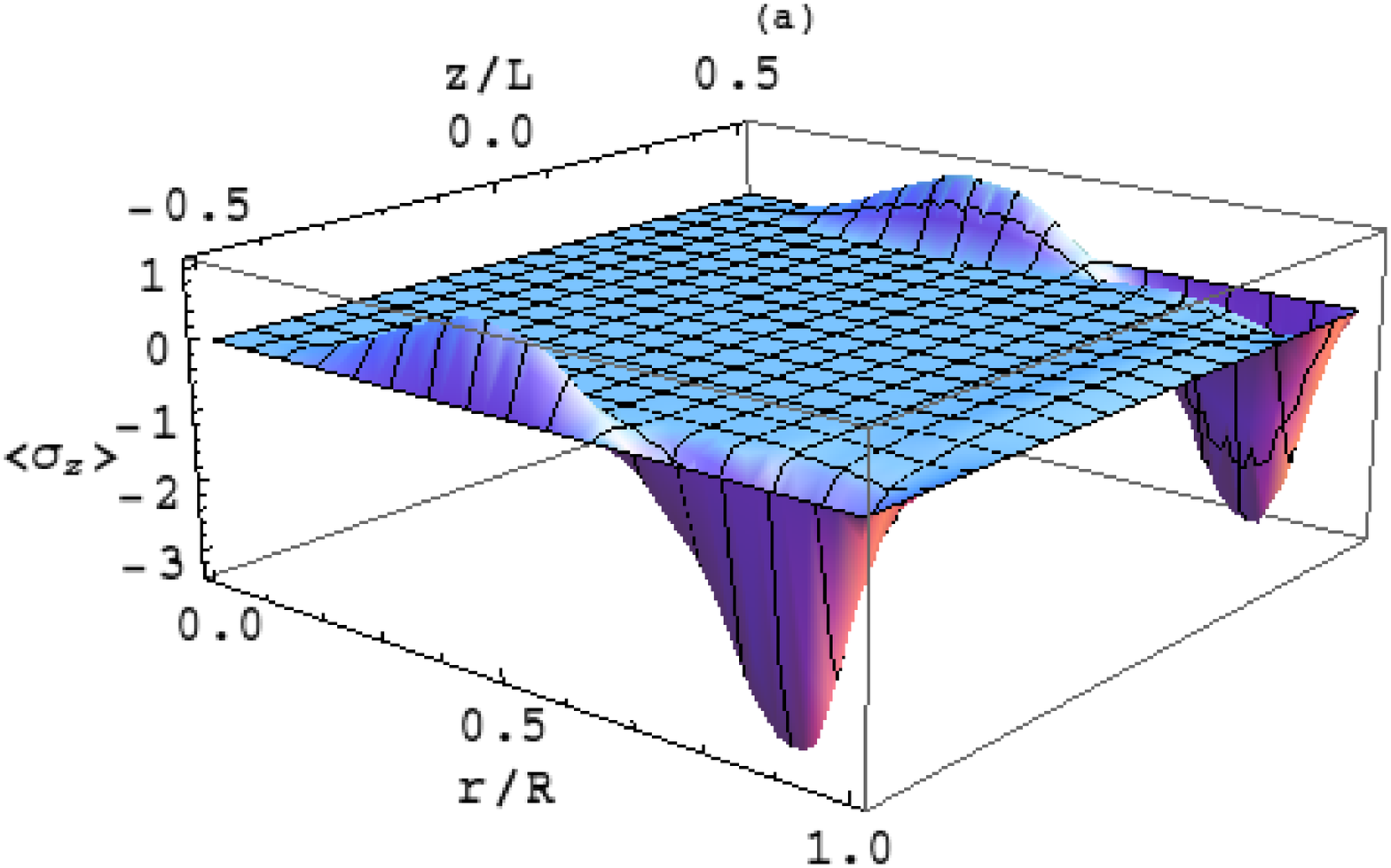}
\includegraphics[width=0.3\textwidth,angle=-90]{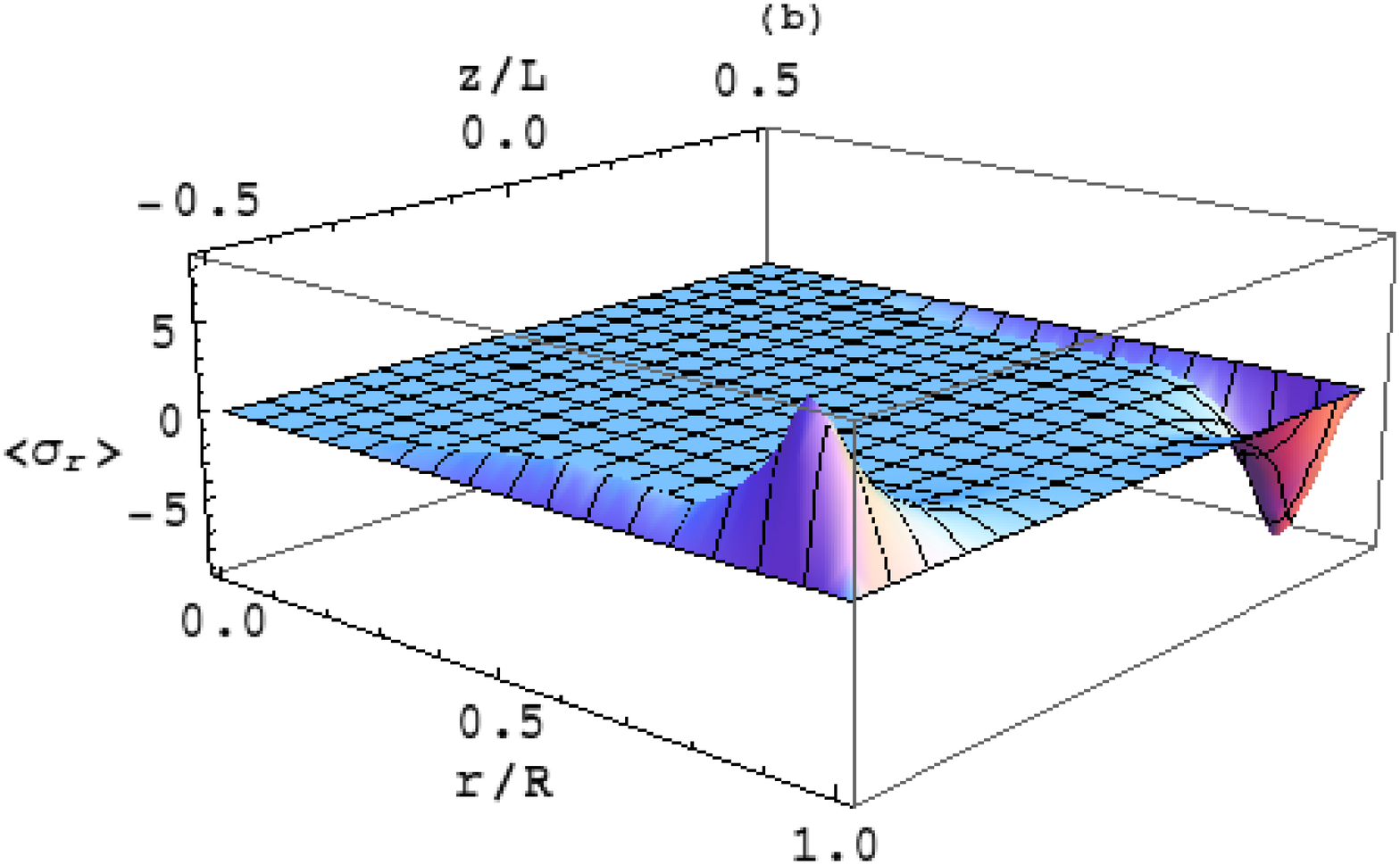}
\caption{\label{fig4}  (Color online) 
Same as Fig.~\ref{fig2} but for the ``subgap'' state with $j=3/2$
indicated by the right arrow in Fig.~\ref{fig1}.}
\end{figure}

For $j>1/2$, our numerical results 
include an almost degenerate pair of subgap states, where
the degeneracy is on top of the Kramers degeneracy.
The charge density is then localized with equal probability near each of the 
two cylinder caps.  A typical example for the spin texture
of such a subgap state is shown (for $j=3/2$) in Fig.~\ref{fig4}.
The out-of-plane spin part is identical on both caps, see Fig.~\ref{fig4}(a),
 but the in-plane (radial) component shown in Fig.~\ref{fig4}(b) 
has opposite direction.  

Comparing our numerical results for parity, charge 
and spin densities, we find that for each eigenstate,
they are linked by a set of general relations.  In particular, for 
the (radially integrated)  densities in the trunk region, we find
\begin{equation} \label{numparity}
\langle \tau_y\rangle(z) \propto - \langle\sigma_r\rangle(z),
\quad   \langle\tau_z\rangle(z)   \propto \langle\rho\rangle(z).
\end{equation}
Note that in the infinite wire case,\cite{egger,paco} the
parity structure is trivial in the sense that (for large $R$)
the only non-zero component is $\langle\tau_z\rangle$,
cf.~Appendix~\ref{appa}.

\section{Surface Dirac fermion theory}
\label{sec3}

In this section, we analyze the finite-length nanowire geometry of 
Sec.~\ref{sec2} within a surface Dirac fermion theory, where
we retain parity in the Hilbert space of the surface Hamiltonian.  
While this is not necessary for an 
infinite cylinder\cite{egger} or for the flat  
2D surface,\cite{zhang2} the discussion at the end
of Sec.~\ref{sec2b} shows that this extension is important here.

\subsection{Infinitely long wire and symmetries}\label{sec3a}

We start with the parity-extended surface Dirac fermion Hamiltonian for an
infinitely long nanowire,\cite{ashv0,egger}
\begin{equation} \label{hd}
H_D = \left[ v_1 \sigma_\phi (-i\partial_z) -
\frac{v_2}{R} \sigma_z \hat J \right] {\cal T},
\end{equation}
where the total angular momentum operator 
$\hat J$ has been defined in Eq.~\eqref{jdef}. 
${\cal T}={\cal T}^\dagger$ acts in parity space 
and is determined below.  
We note that $H_D$ respects all symmetries present in $H_b$ 
 [Eq.~\eqref{bulkham}].
Specifically, azimuthal symmetry implies 
$[ H_D , \hat J] = 0$, and states are classified by half-integer $j$,
\begin{equation}\label{1d}
\psi(\phi,z) = e^{-i \sigma_z \phi/2} \sum_{j\in 
\mathbb{Z}+1/2} e^{i j \phi} \, \psi_j(z),
\end{equation}
with the 1D spinor $\psi_j(z)$.
Time-reversal symmetry implies $[H_D,\Theta]=0$,  
with the time-reversal operator $\Theta = i \sigma_y \tau_0 \, {\cal C}$,
where ${\cal C}$ denotes complex conjugation. Finally,  
Eq.~\eqref{hd} exhibits inversion symmetry, 
$[H_D,{\cal I}]=0$, with the inversion operator\cite{foot1} 
\begin{equation}\label{invert}
{\cal I} = R_z R_\phi \sigma_0 \tau_z.
\end{equation} 
Here, $R_z$ inverts the $z$ coordinate, $z\to -z$, 
and $R_\phi$ shifts $\phi\to \phi+\pi$.
The parity structure in Eq.~\eqref{invert} follows 
from the results of Ref.~\onlinecite{zhang1}.
Evidently, both time-reversal and inversion symmetry are only 
kept intact when choosing
 ${\cal T} \in \{ \tau_0, \tau_z \}$ in Eq.~\eqref{hd}.
We here set ${\cal T} = \tau_z$, as follows
from the analytical derivation of Eq.~\eqref{hd} in Appendix \ref{appa}
as well as from numerical calculations based on the Zhang model
for the infinite nanowire case.\cite{egger} 
 
Using Eq.~\eqref{1d}, we can now switch to a 1D representation for
a given angular momentum ($j$) channel.
For given energy $E$, the 1D spinor $\psi_j$ 
obeys the 1D Dirac equation $H^{(j)}\psi_j=E\psi_j$ with\cite{foot2}
\begin{equation}\label{scheq} 
H^{(j)} = \left( -
 i v_1 \sigma_y  \partial_z -\frac{j v_2 }{R} \sigma_z \right) \tau_z ,
\end{equation}
see also the derivation leading to Eq.~\eqref{finalapp} in Appendix \ref{appa}.
Note that the representation of the Dirac matrices $\hat\gamma_k$
in Eq.~\eqref{finalapp} in terms of products of spin and parity matrices,
see Eq.~\eqref{scheq}, is multi-valued and leads to a double counting
of all surface states derived from Eq.~\eqref{scheq}.  
Nevertheless, it is technically convenient to proceed in this 
representation, since the double counting can be easily circumvented,
see Sec.~\ref{sec3c}.
 
The general solution to the 1D Dirac equation with Hamiltonian $H^{(j)}$
in Eq.~\eqref{scheq} reads 
\begin{equation}\label{psij}
\psi_{j} (z) =  \left( \begin{array}{c} 
A_1\chi_j \\ -i A_2 \sigma_x \chi_j \end{array} \right) e^{i k z}
+  \left( \begin{array}{c} B_1 \chi_j^\ast \\ iB_2\sigma_x
\chi_j^\ast \end{array} \right) e^{-i k z},
\end{equation}
with arbitrary complex coefficients $(A_1,A_2,B_1,B_2)$.
The spinors above are in parity space while 
$\chi_j$ acts in spin space,
\begin{equation}\label{chidef}
\chi_j = \left( \begin{array}{c} \cos \gamma_j \\ i 
\sin \gamma_j \end{array} \right) \equiv  \left( \begin{array}{c}
 \sqrt{\frac12 - \frac{jv_2}{2 R E}} \\
i \ {\rm sgn} (E) \sqrt{\frac12 + \frac{jv_2}{2 R E} }
\end{array}\right).
\end{equation}
In Eq.~\eqref{psij}, the longitudinal momentum $k=k(E)$
follows from $v_1 k = \sqrt{E^2 - ( jv_2 / R)^2}$.
Below we consider energies where $k$ is real and positive. 
For a description of the subgap states discussed
in Sec.~\ref{sec2b}, evanescent modes need to be studied instead.
In the 1D representation, the inversion operator
[Eq.~\eqref{invert}] becomes ${\cal I}\to \tilde{\cal I}$ with 
\begin{equation}\label{1dinv}
 \tilde {\cal I} =  R_z \sigma_z\tau_z.
\end{equation}
Since Eq.~\eqref{scheq} stays invariant under 
inversion, $[H^{(j)},\tilde {\cal I}]=0$,
the eigenfunctions (\ref{psij}) can be classified
as inversion symmetric or antisymmetric ($\sigma=\pm$), 
\begin{equation}\label{calIpsi}
\tilde {\cal I} \psi_{j}^{(\sigma)}(z) = \sigma \psi_{j}^{(\sigma)}(z) .
\end{equation}
{}From Eqs.~(\ref{psij}) and (\ref{calIpsi}), 
after a short calculation, we can therefore infer
relations between the coefficients for given inversion symmetry ($\sigma)$:
\[ 
A_1 = \sigma B_1, \quad A_2 = - \sigma B_2.
\]
A general inversion-symmetric (antisymmetric) state thus takes the form 
\begin{eqnarray}\label{trunk}
\psi_{j}^{(+)} &=& \left( \begin{array}{c}
A_1 \\ -iA_2\sigma_x\end{array}\right) \otimes
\left( \begin{array}{c} \cos \gamma_j \ \cos (kz) \\
- \sin \gamma_j\ \sin (kz) \end{array} \right) ,\\ \nonumber
\psi_{j}^{(-)} &=& \left( \begin{array}{c}
iA_1 \\ A_2\sigma_x\end{array}\right) \otimes
\left( \begin{array}{c} \cos \gamma_j \ \sin (kz) \\
 \sin\gamma_j \ \cos (kz) \end{array} \right) .
\end{eqnarray}
Both are parametrized by two complex numbers 
($A_1$ and $A_2$), where the first (second) spinor refers 
to parity (spin) space.

Next let us take the local expectation value for the
charge, spin, and parity operators in
a given general eigenstate $\psi_{j}^{(\sigma)}$.
First, with $\sin^2(\gamma_{-j})=\cos^2(\gamma_j)$,
the charge density is
\begin{widetext}
\begin{equation}\label{dens}
\langle\rho\rangle(z) = 
\left(\left| A_1 \right|^2 + \left| A_2 \right|^2 \right)
\left[ \cos^2(\gamma_{\sigma j})\cos^2 (k z) +
\sin^2 (\gamma_{\sigma j}) \sin^2 (k z) \right].
\end{equation}
For the spin density, we obtain
\begin{equation}\label{eqf}
\left( \begin{array}{c} \langle\sigma_r\rangle \\
\langle \sigma_\phi\rangle\\ \langle \sigma_z \rangle\end{array}
\right)(z) = \sigma \left(\begin{array}{c}
-\frac12\left( | A_1 |^2 + | A_2|^2 \right)
\sin (2\gamma_j) \sin(2k z) \\
0 \\
 \left( |A_1|^2 -| A_2|^2 \right)
\left[ \cos^2 (\gamma_{\sigma j}) \cos^2 (k z) -
\sin^2 (\gamma_{\sigma j}) \sin^2 (k z)\right]
\end{array}\right).
\end{equation}
Similarly, the parity density is obtained in the form
\begin{equation}\label{HD}
\left( \begin{array}{c} \langle\tau_x\rangle \\
\langle \tau_y\rangle\\ \langle \tau_z \rangle\end{array} \right)(z)
= \left( \begin{array}{c} \sigma \, {\rm Im}( A_1 A_2^\ast )
\sin (2\gamma_j) \sin (2k z) \\ \sigma \, {\rm Re}(A_1 A_2^\ast)
\sin (2\gamma_j) \sin (2k z)  \\ \left( | A_1|^2 - | A_2|^2 \right)
\left[ \cos^2(\gamma_{\sigma j}) \cos^2 (k z) + 
\sin^2 (\gamma_{\sigma j})  \sin^2 (k z) 
\right] \end{array}\right).
\end{equation}
\end{widetext}
At this stage, the above results hold for an \textit{arbitrary}\ 
inversion-symmetric ($\sigma=+$) or 
antisymmetric ($\sigma=-$) state. Remarkably, 
the circumferentially oriented spin density 
$\langle\sigma_\phi\rangle$ always vanishes. This is in accordance 
with our numerical observations, see Eq.~\eqref{numzero}.
The current density along the $z$ direction can be obtained from the
local operator\cite{egger}
\begin{equation}\label{jz}
j_z = v_1 \sigma_\phi \tau_z,
\end{equation}
and therefore vanishes identically for these states as well.\cite{footcur}
This result stays valid for arbitrary inversion-symmetric 
boundary conditions (which do not mix $\psi_j^{(\pm)}$ states).  

In order to reach agreement with the numerical results in
Sec.~\ref{sec2b}, the coefficients $A_{1,2}$ should 
obey the three relations
\begin{equation}\label{auxrel}
{\rm Im} (A_1 A_2^\ast) = 0, \quad 
{\rm Re} ( A_1 A_2^\ast) > 0,\quad
|A_1| \neq |A_2|.
\end{equation}
Indeed, the first relation implies consistency with Eq.~\eqref{numzero}. 
The second relation ensures that $\langle \tau_y \rangle (z) 
\propto - \langle \sigma_r \rangle (z)$, see Eq.~\eqref{numparity}.
The third relation is required to have non-vanishing 
spin-parity densities $\langle \sigma_z\rangle$ and 
 $\langle \tau_z \rangle (z)$. Moreover, notice that then
$\langle \tau_z\rangle(z) \propto \langle\rho\rangle(z)$, in accordance with
Eq.~\eqref{numparity}.  

\subsection{Matching trunk and cap states}
\label{sec3b}

The coefficients $A_{1,2}$ in Eq.~\eqref{trunk} 
as well as the energy spectrum and the corresponding eigenstates can 
now be obtained analytically
by matching the trunk states, see Eq.~(\ref{trunk}) for $|z|<L/2$,
 with cap states at $z=\pm L/2$.
Each cap is described by  a surface Dirac  Hamiltonian of the form
\begin{equation}\label{capham}
H_{\rm cap} = v_2 \left[ -i\left(\partial_r+\frac{1}{2r}\right)\sigma_r +
\frac{\hat J}{r}\sigma_\phi\right]\tau_x.
\end{equation}
The parity matrix $\tau_x$ is  uniquely determined by imposing
time-reversal and inversion symmetry, and also appears in
Eq.~\eqref{bulkham}.
In the angular momentum (1D) representation, for given energy $E$, a 
cap state $\Psi_j^{(\sigma)} (r,z= \zeta L/2)$ [with $\zeta=\pm$ and
inversion symmetry index $\sigma=\pm$] takes the general form 
(for $\sigma=\zeta=-$, this should be multiplied by $-1$)
\begin{equation}\label{cap}
\Psi_j^{(\sigma)}(r,\zeta L/2)  = 
 C_\zeta \left(\begin{array}{c} 1 \\ 1 \end{array}\right)
\otimes \xi_j(r) + C_{-\zeta} 
\left(\begin{array}{c} 1 \\ -1 \end{array}\right) 
\otimes \xi_j^\ast(r) 
\end{equation}
with complex coefficients $C_\pm$ and the spin spinor
\begin{equation}
\xi_j(r) = \left( \begin{array}{c} J_{j-1/2}( |E| r/v_2) \\ 
i \, {\rm sgn}(E) J_{j+1/2}(| E| r/v_2) \end{array} \right) .
\end{equation}
The other spinors in Eq.~\eqref{cap} are in parity space. 

Matching the trunk states [Eq.~\eqref{trunk}] and the cap states
[Eq.~\eqref{cap}] for given $j$ and $\sigma$ by continuity 
at $z=\pm L/2$ and $r=R$, we 
obtain four linear equations for the four 
coefficients $(A_1, A_2, C_+, C_-)$.  A nontrivial solution follows
when the corresponding determinant vanishes, yielding the condition 
($u=|E|R/v_2$) 
\begin{equation}\label{energyc}
\left( J^2_{j-1/2}(u)+J^2_{j+1/2}(u)\right) \sin(2\gamma_j) \sin(kL) = 0.
\end{equation}
For real-valued $k\ge 0$, this equation can only be satisfied when $\sin(kL)=0$.
This implies the standard longitudinal momentum quantization condition
$k_n(E) = n \pi/ L$ with $n\in \mathbb{N}_0$.
The corresponding eigenenergies then follow from the 
bulk dispersion relation in Eq.~\eqref{bulkdisp},
\begin{equation}\label{mom}
E_{j,n,\pm} = \pm \sqrt{(  \pi n v_1 / L )^2 + 
\left(j v_2 / R\right)^2 }.
\end{equation}
For a given level,  the wave function amplitudes $A_{1,2}$ and $C_\pm$ 
then satisfy three conditions  plus the overall normalization constraint.
With $p =\sigma (-)^n=\pm$ and the above definition of $u$,
we have $C_+ = pC_-$ and the relation
\begin{equation}\label{wave1}
J_{j+p/2}(u) A_1 + p \ {\rm sgn}(E) \ J_{j-p/2}(u) A_2 = 0.
\end{equation}
Moreover, for $p=+$ the third condition reads
\begin{equation}\label{wave2}
  \cos(\gamma_j)  A_1 - 2  (-i)^n  J_{j-1/2}(u) C_+ =0 ,
\end{equation}
while for $p=-$ this instead becomes
\[
  \sin(\gamma_j)  A_1 - 2 (-i)^n \ {\rm sgn}(E) \ J_{j+1/2}(u) C_+ =0 .
\]
These relations determine all possible wavefunctions for the 
closed cylinder surface.

Of course, since we considered only solutions of Eq.~\eqref{energyc} with 
real $k$, subgap states were not captured. 
However, within the linear-in-${\bm k}$ approximation underlying
the approach here, we find that there are no subgap states at all, 
i.e., the corresponding matching problem with evanescent trunk modes 
does not permit a nontrivial solution.  The numerical approach of 
Sec.~\ref{sec2} also indicates that in order to obtain subgap states, it is 
necessary to include higher-order terms (in ${\bm k}$) breaking electron-hole 
symmetry in the Hamiltonian.

\subsection{Spectrum and eigenstates of the dot}
\label{sec3c}

Using the above explicit solution for the wavefunction of the
complete cylinder, we obtain the coordinate dependence of
all densities of interest.  We here describe the results
for the trunk region only.  First of all, we recover 
\[
\langle \sigma_\phi\rangle= \langle\tau_x\rangle=0, 
\] 
see Eq.~\eqref{numzero}.  Moreover, the relations 
\[
\langle \tau_y\rangle \propto - \langle\sigma_r\rangle,
\quad \langle\tau_z\rangle  \propto \langle\rho\rangle 
\]
are also reproduced, see Eq.~\eqref{numparity}.  
Specifically, the dot eigenstates have the energy $E\equiv
E_{j,n,\pm}$ specified
in Eq.~\eqref{mom}.  We show below that these energies are not degenerate,
i.e., only one specific inversion parity $\sigma$ given by 
\begin{equation}\label{selectionrule}
\sigma\equiv \sigma_{j,n,\pm} = \mp (-1)^n {\rm sgn}(j) 
\end{equation}
will be physically realized.  With $k\equiv \pi n/L\ge 0$, 
we find for the charge and spin densities  ($u=|E|R/v_2$)
\begin{eqnarray} \nonumber
\langle\rho\rangle(z) & \propto &  1 - {\rm sgn}(E) \frac{ j \sigma}{u}
 \cos(2kz)  , \\ \label{depen}
\langle\sigma_r\rangle (z) & \propto & \sqrt{1-j^2/u^2} \ \sin(2kz),\\
\nonumber \langle \sigma_z \rangle (z) & 
\propto & \cos(2kz) - \frac{j \sigma}{u} {\rm sgn}(E).
\end{eqnarray}
These results are in good agreement with the numerical results for 
the spin texture obtained for the Zhang model in Sec.~\ref{sec2}.
In particular, they show explicitly that spin-surface
locking is broken.  

Interestingly, for each total angular momentum $j$, there are two 
zero-momentum states corresponding to conduction and valence
band, respectively. Their inversion symmetry properties are determined by 
\begin{equation}\label{zeromomcond}
\sigma = - {\rm sgn}(j E_{j,n=0,\pm}),
\end{equation}
since for states with the opposite value of $\sigma$, all densities 
in Eq.~\eqref{depen} vanish.
For the physically allowed $k=0$ state with $\sigma$ in 
Eq.~\eqref{zeromomcond},  from Eq.~\eqref{depen} we instead find spatially
\textit{uniform} densities $\langle\rho\rangle(z)$, 
$\langle \tau_z \rangle (z)$, and $\langle \sigma_z \rangle (z)$,
while all remaining spin or parity density components vanish.

We now compare the densities in Eq.~\eqref{depen} to the numerical
results in Sec.~\ref{sec2}, and also address the double counting 
problem mentioned in Sec.~\ref{sec3a}.
Eq.~\eqref{wave1} shows that indeed ${\rm Im}(A_1 A_2^\ast) =0$, 
in accordance with our numerical results in Sec.~\ref{sec2b}, see
Eq.~\eqref{auxrel}.  Also the relation $|A_1|\ne |A_2|$
in Eq.~\eqref{auxrel} is evidently satisfied.
However, not all the states can  be realized  physically,
as is clear by comparing to the condition ${\rm Re}(A_1 A_2^\ast)>0$ in 
Eq.~\eqref{auxrel} found numerically in Sec.~\ref{sec2}.
In order to understand this restriction, we note that the operator 
$\Xi=\sigma_z\tau_z$ commutes both 
with the cap Hamiltonian [Eq.~\eqref{capham}]
and with the inversion operator $\tilde {\cal I}$ [Eq.~\eqref{1dinv}].
This implies by continuity that trunk states 
[Eq.~\eqref{trunk}] at the end points ($z=\zeta L/2$
with $\zeta=\pm$) are eigenstates of $\Xi$ as well,
\[
\Xi \psi_{j,n}^{(\sigma)} (\zeta L/2) =  p \psi_{j,n}^{(\sigma)}
(\zeta L/2),
\]
where the eigenvalues $p=\sigma (-1)^n$ follow from 
Eq.~\eqref{trunk} and the definition of $\Xi$.  For $n=0$,
however, Eq.~\eqref{zeromomcond} implies 
that only the eigenvalue $p=-{\rm sgn}(jE_{j,0,\pm})$ 
is physically realized. By continuity, this value must also apply
for the full Hilbert space of conduction or valence surface bands.
We therefore obtain the ``selection rule'' in Eq.~\eqref{selectionrule}
restricting the Hilbert space of allowed states. 
This explains the condition ${\rm Re}(A_1 A_2^\ast)>0$ and
resolves the double-counting problem. 
We mention in passing that the latter problem is 
automatically avoided when retaining terms of order ${\bm k}^2$ in
the Hamiltonian, where the spin-parity eigenstates with $p=\pm$ have
different energy.  In the Dirac theory, this implies a ``spontaneously
broken symmetry'' encoded by Eq.~\eqref{selectionrule}.

\subsection{Effective boundary conditions}
\label{sec3d}

It is also possible to derive the results in Sec.~\ref{sec3c}
without explicit construction of the cap states.  To that
end, let us briefly consider a class of general boundary conditions 
at the cylinder ends, $z=\zeta L/2$ with $\zeta=\pm$, by 
imposing the local gauge constraints
\begin{equation}\label{bc}
\psi(\phi,\zeta L/2) = \Lambda_\zeta \, \psi(\phi,\zeta L/2),
\end{equation}
where $\Lambda_\zeta = \Lambda_\zeta^{-1} = \Lambda_\zeta^\dagger$.
The spin-parity structure of $\Lambda_\zeta$ can be determined by requiring
 time-reversal invariance, $[ \Lambda_\zeta , \Theta] = 0$, and
invariance under inversion, $[\Lambda_\zeta , {\cal I}] = 0$.
In addition, we require the boundary operator to commute with $H_D$,
which is a natural assumption for closed surfaces.
 As a result, with arbitrary angles $\eta_\pm$, we find
\[
\Lambda_\zeta  = \sin (\eta_\zeta ) \sigma_0 \tau_z +
\cos (\eta_\zeta) \sigma_r\tau_y.
\]
Passing to the 1D representation, i.e., for given half-integer $j$, 
these constraints read
\begin{eqnarray}\label{bc2}
\psi_j(\zeta  L/2) &=& \tilde\Lambda_\zeta  \psi_j(\zeta L/2),\\ \nonumber
\tilde\Lambda_\zeta  &=& \sin (\eta_\zeta) \sigma_0  \tau_z +
\cos (\eta_\zeta) \sigma_x  \tau_y .
\end{eqnarray}
Applying the inversion operator $\tilde {\cal I}$, see Eq.~\eqref{1dinv},
to the boundary condition (\ref{bc2}), we find 
$\tilde\Lambda_+= \tilde\Lambda_-$  and hence
$\eta_\pm\equiv \eta$. Only then will inversion symmetry be preserved
for the confined states.
The parameter $\eta$ (with $0\le \eta< \pi$) 
cannot be fixed by symmetry considerations alone but 
depends on the physical boundary condition imposed at the ends, i.e.,
the boundary matrix $\tilde\Lambda$ 
effectively encodes the  matching of trunk states with 
cap states. Contrary to the commonly employed boundary 
conditions,\cite{falko,akhmerov} the operator
$\tilde\Lambda$
\textit{commutes}\ with the current operator $j_z$, see  Eq.~\eqref{jz},
while the anticommutator is always nonzero. 
Since the boundary conditions are invariant with respect to
 inversion, they do not mix the states (\ref{trunk})  
with opposite inversion parity $\sigma$.
Using the boundary condition (\ref{bc2}), some algebra yields for both
solutions and for arbitrary energy $E$ the condition
\begin{equation}\label{AA}
A_1 + \frac{\cos \eta}{1-\sin \eta} A_2 = 0.
\end{equation}
Comparing this to Eq.~\eqref{wave1}, the energy-dependent 
angle $\eta$ can be explicitly related to the above wavefunction
matching procedure, and the subsequent results in Sec.~\ref{sec3c}
can be obtained under a purely 1D description of the trunk states alone.

\section{Microscopic tight-binding approach}\label{sec4}

A simple microscopic model for a strong TI was previously proposed by
Fu, Kane, and Mele.\cite{fu-kane-mele-07} The model consists of a 
single-band tight-binding model on a diamond lattice and 
includes spin-orbit couplings.  With lattice fermion operators
$c_i$ (spin is kept implicit), this Hamiltonian has the form
\begin{equation}\label{tb}
H_{\rm tb} = \sum_{\langle i,j\rangle} t_{ij} c_i^\dagger c_j^{}
+ \frac{ 4i\lambda_{\rm so}}{a^2 }
\sum_{\langle\langle i,j\rangle\rangle}
c_i^\dagger \left( {\bm \sigma}\cdot  \left
[ {\bm d}_{ij}^1 \times {\bm d}_{ij}^2 \right] \right) c_j^{},
\end{equation}
where $a$ is the cubic lattice constant, $t_{ij}$ are hopping parameters 
connecting nearest neighbors, 
and the last term describes spin-orbit coupling
of strength $\lambda_{\rm so}$ through a second-neighbor hopping
between sites $i,j$, which depends on the two nearest-neighbor vectors 
${\bm d}^{1,2}$ connecting those two sites. 
In order to generate a full gap in the bulk spectrum, a 
distortion $t_{ij} \rightarrow t + \delta t$ is introduced for 
${\bm d}_{ij}$ along the (111) direction.\cite{fu-kane-mele-07}

To define the nanowire, we proceed as in Ref.~\onlinecite{egger} by
selecting the growth direction $\hat{e}_z$ along the (111) axis and 
keeping all sites within a given radius $R$. 
The unit cell of the infinite nanowire thus defined contains six planes of 
sites corresponding to the three stackings of the two fcc sublattices of the 
diamond lattice, and has the period $d_{\rm cell} = \sqrt{3} a$.
Although this model is not completely equivalent to the 
one of Eq.~(\ref{bulkham}), quantitatively similar behavior of the 
surface gap $\Delta_s$ as 
a function of $R$ has been obtained\cite{egger} 
by setting $a=2.8$~nm and $-2t + \delta t = M_0$.
The finite dot geometry is then defined by setting the 
length of the nanowire along the (111) direction to a given value $L$. 
To maintain the aspect ratio of the cylindrical dot 
studied numerically in Sec.~\ref{sec2},
we set $R=3a$ and $L=4\sqrt{3}a$. This corresponds to a cluster
of 1592 sites, which is a small enough size to keep a reasonable 
computational cost of the calculations 
while still allowing for a meaningful comparison of the spin texture
 with the results of Secs.~\ref{sec2} and \ref{sec3}.

\begin{figure}
\includegraphics[width=0.55\textwidth]{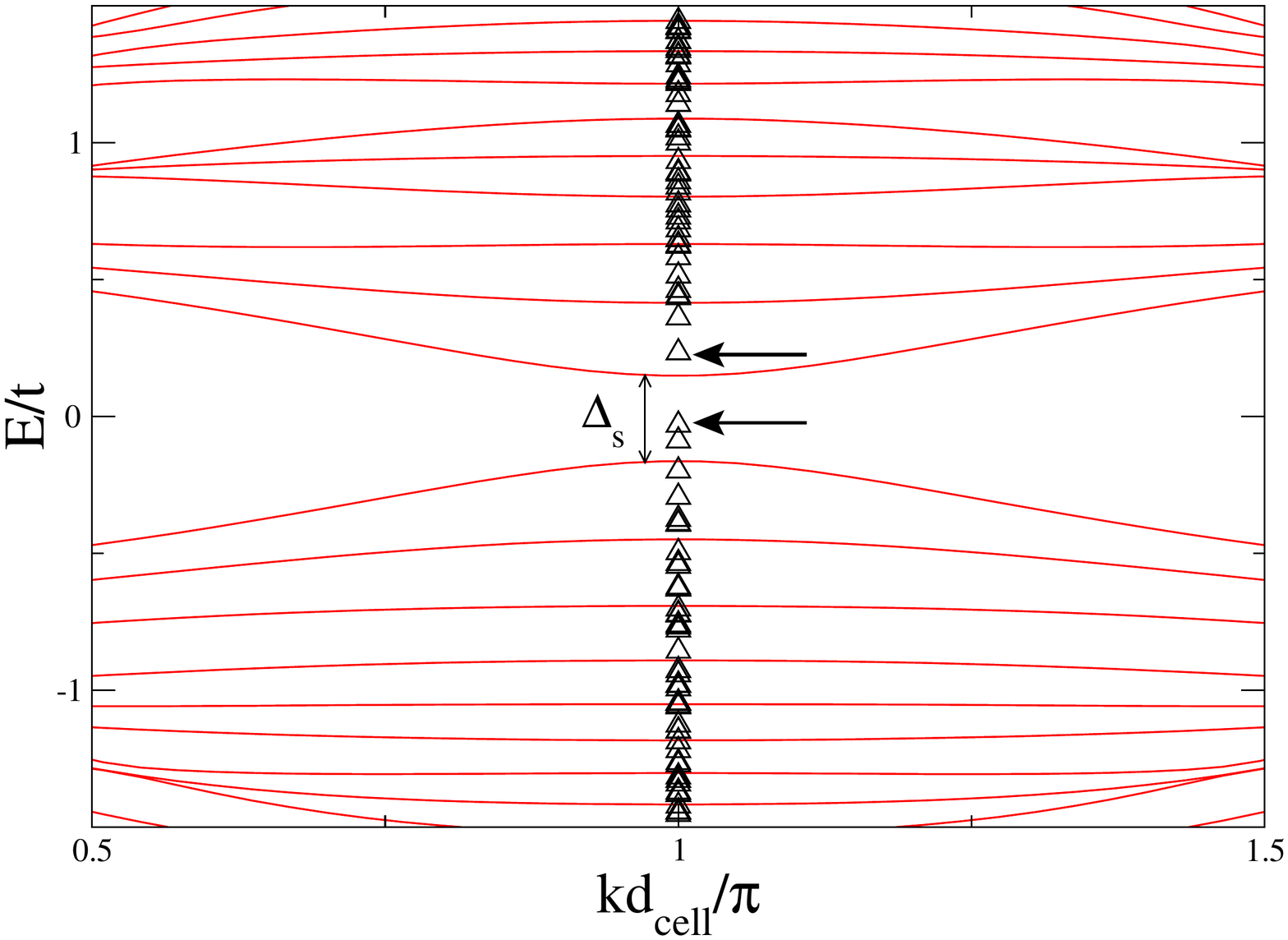}
\caption{\label{fig5}  (Color online) 
Band structure of the infinite TI nanowire (red solid curves) 
and energy levels (denoted by open black triangles) of a
finite dot. These results follow from the tight-binding model [Eq.~(\ref{tb})]
with wire axis along the (111) direction. 
The dot length is set to $L=4\sqrt{3}a$, and the radius is $R=3a$. 
The arrows indicate the two levels whose spin texture 
is analyzed in Figs.~\ref{fig6}
and \ref{fig7}, respectively.}
\end{figure}

The band structure of the infinite nanowire and the energy 
levels in the finite dot geometry are shown in Fig.~\ref{fig5}, 
where we again focus on states energetically inside the bulk gap.
As expected, both the bands of the infinite wire and all dot levels 
are twofold Kramers degenerate. For the chosen parameter set, 
we find two subgap states appearing inside the surface gap $\Delta_s$. 
(Since there is no full rotational symmetry anymore, we cannot
classify states by $j$ here.)

\begin{figure}
\includegraphics[width=0.4\textwidth]{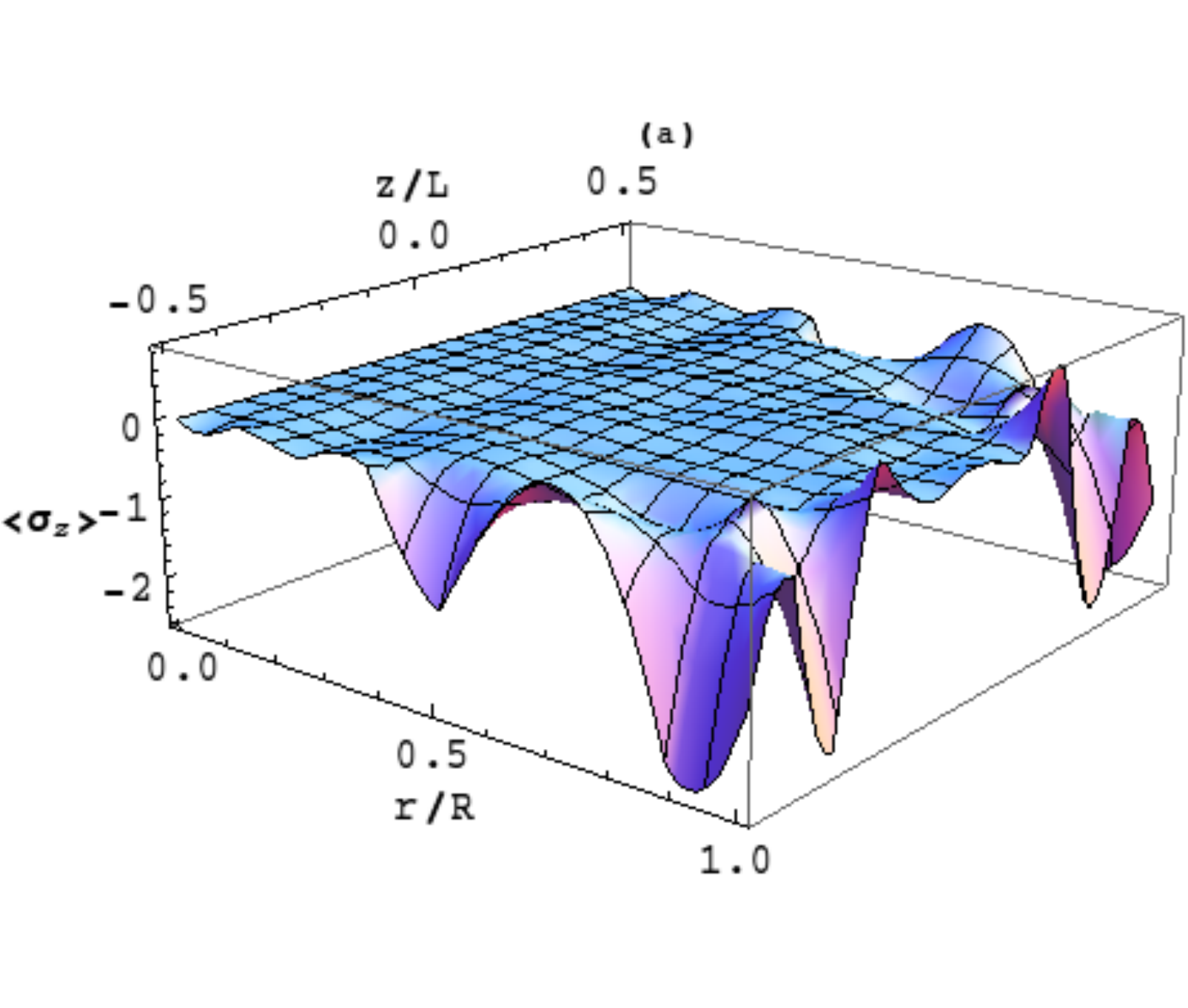}
\includegraphics[width=0.4\textwidth]{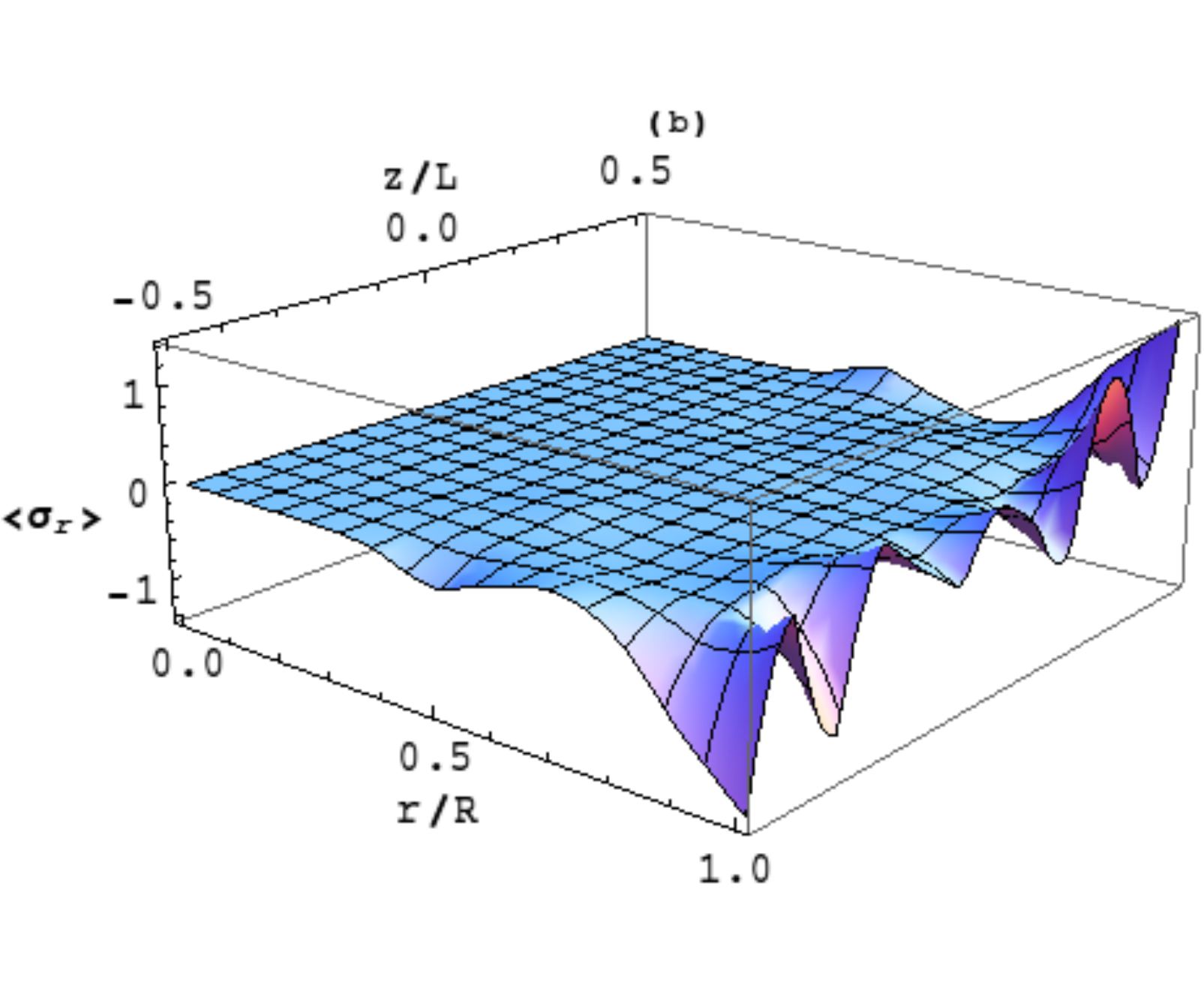}
\caption{\label{fig6} (Color online)  Spin texture in the $rz$ plane
obtained from the tight-binding model [Eq.~(\ref{tb})]. The 
spin densities $\langle\sigma_z\rangle$ [(a)] and
$\langle\sigma_r\rangle$ [(b)] are shown for the subgap state 
indicated by the lower arrow in Fig.~\ref{fig5}.}
\end{figure}

\begin{figure}
\includegraphics[width=0.4\textwidth]{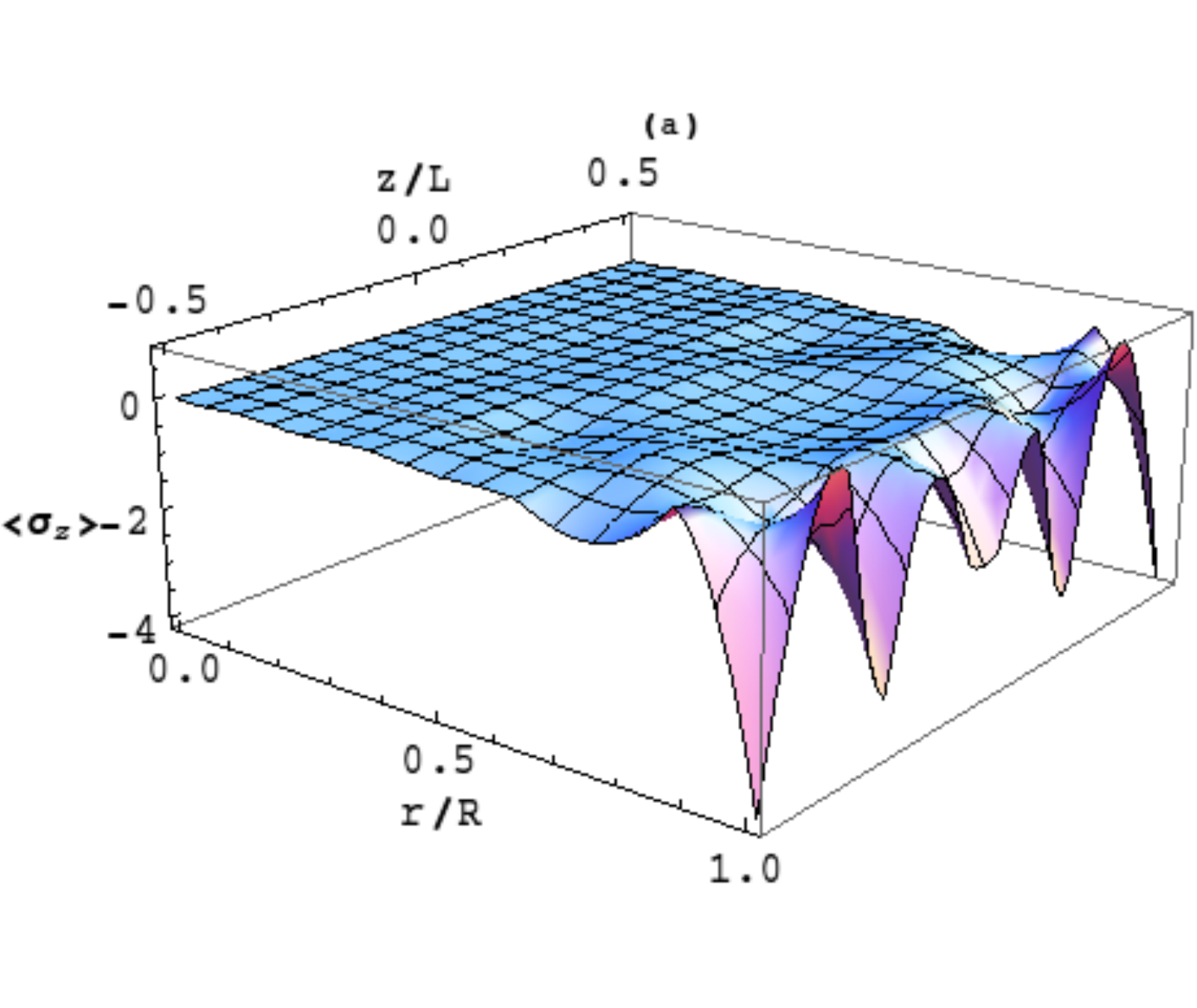}
\includegraphics[width=0.4\textwidth]{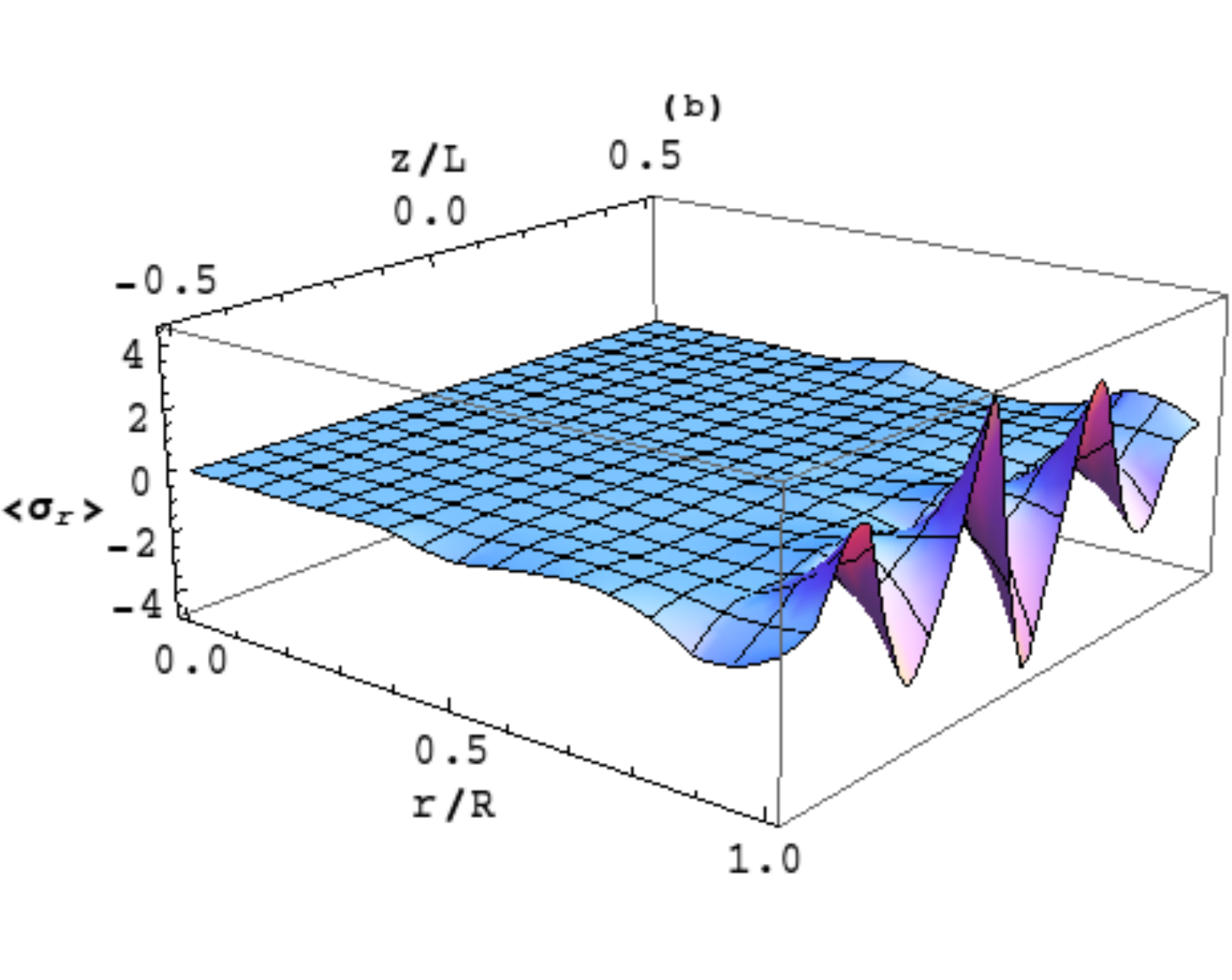}
\caption{\label{fig7}  (Color online) Same as Fig.~\ref{fig6} but for the
state indicated by the higher arrow in Fig.~\ref{fig5}.}
\end{figure}

To get an idea of the spin texture of the dot states, we now
focus on the two states indicated by arrows in Fig.~\ref{fig5}.
The lower arrow corresponds to a subgap state, and 
the second one refers to the lowest-lying state within the 
conduction band of the infinite wire. 
The corresponding spin densities are shown in Figs.~\ref{fig6} and 
\ref{fig7}.  
As in the effective low-energy theory and the 
surface Dirac fermion description, we again observe  
$\langle\sigma_{\phi}\rangle=0$, and 
therefore only $\langle\sigma_z\rangle$ and $\langle\sigma_r\rangle$ are shown.
Note that in the infinite wire case, one would instead 
find $\langle \sigma_r\rangle=0$
consistent with spin-surface locking.\cite{egger}

The subgap states have a charge density mostly localized 
near the caps of the cylindrical dot, again with out-of-plane (in-plane) spin 
components that are identical (oppositely directed) on both sides, 
reproducing the results of Secs.~\ref{sec2} and \ref{sec3}.
However, unlike the continuous model, the tight-binding model 
predicts a spin texture with a superimposed atomic-scale 
oscillation. This oscillation stems from the finite $k=\pi/d_{\rm cell}$ 
value at the Dirac point in this model.

On the other hand, for the lowest-lying state within the conduction band,
corresponding to the ``zero-momentum'' state in
Secs.~\ref{sec2} and \ref{sec3}, Figure \ref{fig7} shows that the
density is largest along the cylinder trunk. The spin is predominantly 
oriented along the negative $z$-direction, but with a finite 
oscillatory component in the radial direction that breaks
spin-surface locking.  

To conclude, even though the results of the tight-binding model are not fully 
equivalent to the ones in Secs.~\ref{sec2} and \ref{sec3}, we find that the
main properties of the subgap and lowest conduction band states are
reproduced.

\section{Conclusions}
\label{sec5}

In this paper we have studied the band structure of a quantum dot
made of a strong topological insulator using three different approaches,
namely the low-energy theory of Zhang 
\textit{et al.},\cite{zhang1,zhang2} an effective
surface Dirac fermion theory, and numerical calculations for a 
tight-binding model on a diamond lattice with strong spin-orbit couplings.
The considered geometry, with flat caps terminating 
a finite-length cylindrical nanowire, is characterized 
by sharp edges where the cylinder trunk and caps meet.  Such edges 
are also present in typical ``mesoscopic'' TI devices studied 
experimentally.\cite{peng,cha,zuev,tang}
All three approaches show that spin-surface locking is generally violated
due to presence of these edges.  As also found in a recent
\textit{ab initio} study,\cite{moon}
a finite reflection probability for Dirac fermions in each
part results  when two surfaces are patched together.
In our case, we have a Fabry-Perot-like setup where 
standing waves can build up. The resulting spin density then exhibits
spatial oscillations reminiscent of a spin density wave state.
The spin direction of the oscillatory parts points out of 
the surface while  non-oscillatory spin density contributions
stay locked to the surface.

The spectrum of such a quantum dot shows several
surprising features. First, when starting from the band structure
of the infinitely long wire [Eq.~\eqref{bulkdisp}], imposing the usual 
longitudinal quantization condition, $k_n=n\pi /L$,  here allows for
a nontrivial eigenstate with $k=n=0$.
In such a zero-momentum state the charge and spin densities along
the trunk are basically homogeneous.  Second, albeit the surface bands 
of an infinite nanowire exhibit a gap, the finite-length nanowire dot 
has subgap states when electron-hole symmetry is broken.  
The wavefunction of such a subgap state is localized on 
both caps simultaneously.  The obtained energy spectrum and 
corresponding spin textures are important ingredients
for a theory of mesoscopic transport through TI dots.
In general, we also expect Coulomb interactions to be relevant,
in particular charging effects should be visible.  
We plan to address these questions in the future.
Moreover, extensions of the theory to include 
an applied magnetic field, where the typically large and 
anisotropic Land{\'e} factor\cite{zhang2} 
implies that the Zeeman field is crucial,
are also left for future work.

\acknowledgments
We acknowledge financial support by the SFB Transregio 12 of the DFG
and by the Spanish MICINN under contract FIS2008-04209.

\appendix
\section{Surface Hamiltonian for infinite nanowire} \label{appa}

\begin{table}[t!]
\begin{tabular}{|c||c|c|c|c|}
\hline
 & $\tau_0$ & $\tau_x$ & $\tau_y$ & $\tau_z$ \\
\hline
\hline
$\sigma_0$ & $\hat\gamma_0$ & $0$ & $\hat\gamma_x$ & $0$ \\
\hline
$\sigma_x$ & $-\hat \gamma_x$ & $0$ & $-\hat \gamma_0$ & $0$ \\
\hline
$\sigma_y$ & $0$ & $\hat\gamma_z$ & $0$ & $\hat\gamma_y$ \\
\hline
$\sigma_z$ & $0$ & $\hat\gamma_y$ & $0$ & $-\hat\gamma_z$ \\
\hline
\end{tabular}
\caption{\label{tab1} Representation of spin-parity matrices in terms of Pauli matrices $\hat\gamma_i$ acting in surface-state subspace.}
\end{table}

Here we analytically derive Eq.~(\ref{hd})
for an infinite nanowire using the gap-inversion model of 
Ref.~\onlinecite{paco}.  Our starting point is Eq.~\eqref{bulkham}
to linear order in ${\bm k}$ (we put $C_0=0$ here), 
\begin{equation}\label{app1}
H_b = M\sigma_0 \tau_z + \left[ v_1 k \sigma_z + 
v_2 ( k_x \sigma_x + k_y \sigma_y) \right] \tau_x .
\end{equation}
Following Ref.~\onlinecite{paco}, we assume that the 
gap parameter changes sign at $r = R$,
i.e.,  $M(r) = M_0 \ {\rm sgn}(R-r)$ with $M_0>0$.
For $r<R$ ($r>R$), the material is then
in the topologically nontrivial (trivial) phase.
In the 1D representation, for given $j$ and $k$, solutions to the Dirac
equation take the form $e^{i k z} \
 e^{-i \sigma_z \phi/2} \sum_j e^{i j \phi} \, \psi_j(r)$,
with the 1D radial Dirac equation
\[
\left( H_0^{(j)} + v_1 k \sigma_z \tau_x \right) \psi_j = E \psi_j .
\]
Here $k=0$ corresponds to 
\[
H_0^{(j)} = M(r) \sigma_0 \tau_z +  v_2 
\left[ -i \left( \partial_r + \frac{1}{2 r} \right) \sigma_x +
\frac{j}{r}  \sigma_y \right] \tau_x .
\]
For $k=0$, there are two solutions 
\[
\psi_j^{(s=\pm)}(r) = \left( \begin{array}{c} i s \cos \gamma \\ 
-{\rm sgn}(M)  \sin \gamma \end{array} \right)
\otimes \left( \begin{array}{c}
I_{j-1/2}(\kappa r) \\ -s I_{j+1/2}(\kappa r)
\end{array} \right) ,
\]
where the first (second) spinor refers to parity (spin) space,
$\sin \gamma = \sqrt{\frac12 -\frac {E} {2 M}}$,
$\cos \gamma = \sqrt{\frac12 +\frac {E} {2 M}}$, and
$v_2\kappa= \sqrt{M_0^2 - E^2}.$ 
For $r>R$, we have to replace 
$I_{j-1/2}\to K_{j-1/2}$ and $I_{j+1/2}\to -K_{j+1/2}$ 
in the spin part, where $I_\nu$ and $K_\nu$ are modified Bessel functions.
The general solution is then given by
\begin{equation}\label{apsi}
\psi_j(r) = \left\{ \begin{array}{cc}
 \alpha_1 \psi_j^{(+)}(r) + \beta_1 \psi_j^{(-)}(r), & r<R, \\
\alpha_2 \psi_j^{(+)}(r) + \beta_2 \psi_j^{(-)}(r), & r>R, \end{array}\right. 
\end{equation}
where the coefficients $\alpha_{1,2}$ and $\beta_{1,2}$ are obtained
by requiring continuity of the wave function at $r = R$.
This results in a linear system of equations for the coefficients, 
which has a nontrivial solution under the condition
\begin{eqnarray*}
&& \left(\cos^2(\gamma) I_{j-1/2}K_{j+1/2} - \sin^2(\gamma)
 I_{j+1/2} K_{j-1/2} \right ) \\ &\times &  
\left(\sin^2(\gamma) I_{j-1/2}K_{j+1/2} - \cos^2(\gamma)
 I_{j+1/2} K_{j-1/2} \right )=0,
\end{eqnarray*}
where all Bessel functions have the argument $\kappa R$ and
$\gamma$ is evaluated for $M=M_0$.
Assuming $\kappa R \gg 1$ and using the asymptotic form of the Bessel functions,
we find $\cos(2\gamma)\simeq \pm j/(\kappa R)$ and 
thus $E_{j,\pm}(k=0) = \pm j v_2/ R$.
The corresponding $k=0$ wavefunctions, $\psi_{j,\pm}(r)$, are given by
Eq.~\eqref{apsi} with $\beta_1=\pm \alpha_1$, $\beta_2=\pm\alpha_2$, and
\[
\alpha_1  = \sqrt{\frac{\pi}{2}}  \kappa e^{- \kappa R},\quad
\alpha_2 = \mp \frac{\kappa}{\sqrt{2 \pi}}  e^{\kappa R}.
\]
For small $k\neq 0$, the effective surface Hamiltonian, Eq.~(\ref{hd}),
is obtained by projecting Eq.~\eqref{app1}
onto the subspace spanned by the above $k=0$ states.
We thereby find 
\begin{equation}\label{finalapp}
H^{(j)} = \frac{ j v_2 }{ R}  \hat\gamma_z + v_1 k \hat\gamma_y ,
\end{equation}
where  the $\hat\gamma_i$ are Pauli matrices in the zero-momentum subspace.
In this way, all combinations of spin-parity matrices, $\sigma_i \tau_j$,
can be represented in the truncated basis. For $\kappa R \gg 1$, we obtain
the results in Table~\ref{tab1}. Of course, this representation is not 
single-valued, i.e., there is no one-to-one 
correspondence between the $\sigma_i\tau_j$ and the $\hat\gamma_k$ matrices. 
In particular, the spin operators $\sigma_{y,z}$ 
and the parity operator $\tau_z$ are entangled, since
$\sigma_{y,z}\tau_0 = 0$ and $\sigma_0 \tau_z = 0$.
Taking $\hat\gamma_y = \sigma_y \tau_z$ and $\hat\gamma_z 
= -\sigma_z \tau_z$, cf.~the last column of Table \ref{tab1},
one gets the surface Hamiltonian quoted in the main text,
see Eq.~\eqref{scheq}.
Because this representation is multi-valued, the replacement of 
$\hat\gamma_{y,z}$ by products of spin and parity matrices 
causes a double-counting of states if used naively. This double-counting
problem is, however, not severe and can be circumvented, as we
discuss in Sec.~\ref{sec3c} in the main text.


\begin{thebibliography}{10}
\bibitem{malkova}
We note that a theory for quantum dots based on the 
2D semiconductors HgTe and HgS (which
may have topological phases) was recently put forward  
by N. Malkova and G.W. Bryant, Phys. Rev. B {\bf 82}, 155314 (2010).
\bibitem{foot2}
In the 1D representation \eqref{1d}, the cylindrical Pauli 
matrices $\sigma_{r, \phi}$ in Eq.~\eqref{cylpaul} are
replaced by $\sigma_{x,y}$.
\bibitem{footcur}
For current-carrying states, one has to 
impose boundary conditions breaking the inversion symmetry.
\end{thebibliography}
\end{document}